\documentclass[prl,reprint,amsmath,amssymb,superscriptaddress,preprintnumbers]{revtex4-1}
\usepackage{amssymb}
\usepackage{graphicx}
\usepackage{longtable}

\setlongtables

\newcommand{\networkx}{{\tt networkx}}
\newcommand{\graphtool}{{\tt graph-tool}}

\begin{document}

\title{Data mining the EXFOR database using network theory}

\author{J.A. Hirdt}
\affiliation{Department of Mathematics and Computer Science, St. Joseph's College, Patchogue, NY 11772, USA}

\author{D.A. Brown}
\email[Corresponding author: ]{dbrown@bnl.gov}
\affiliation{National Nuclear Data Center, Brookhaven National Laboratory, Upton, NY 11973-5000, USA}

\date{\today} 

\begin{abstract}
{The EXFOR database contains the largest collection of experimental nuclear reaction data available as well as the data's bibliographic information and experimental details. We created an undirected graph from the EXFOR datasets with graph nodes representing single observables and graph links representing the various types of connections  between these observables.  This graph is an abstract representation of the connections in EXFOR, similar to graphs of social networks, authorship networks, etc.  By analyzing this abstract graph, we are able to address very specific questions such as 1) what observables are being used as reference measurements by the experimental nuclear science community?  2) are these observables given the attention needed by various nuclear data evaluation projects?  3) are there classes of observables that are not connected to these reference measurements?  In addressing these questions, we propose several (mostly cross section) observables that should be evaluated and made into reaction reference standards.
}
\end{abstract}

\preprint{BNL-103517-2013-JA}

\maketitle

\section{Introduction}

In the early 1950's, Brookhaven National Laboratory began compiling and archiving nuclear reaction experimental data in the SCISRS database \cite{EXFORHistory}.  Over the years, this project have grown and evolved into the EXFOR project \cite{EXFOR}.  EXFOR is an international experimental nuclear data collection and dissemination project now led by the IAEA nuclear data section, coordinating the experimental nuclear data compilation and archival work of several nations.

The EXFOR nuclear experimental database provides the data which underpins nearly all evaluated neutron and charged particle evaluations in ENDF-formatted nuclear data libraries (e.g. ENDF/B, JEFF, JENDL, ...).  Therefore, EXFOR is in many ways the ``mother library'' which leads to the nuclear data used in all applications in nuclear power, security, nuclear medicine, etc.  The EXFOR database includes a complete compilation of experimental neutron-induced, a selected compilation of charged-particle-induced, a selected compilation of photon-induced, and assorted high-energy and heavy-ion reaction data. The EXFOR library is the most comprehensive collection of experimental nuclear data available. Therefore, it is the best place to look for an overview of what the applied and basic experimental nuclear science communities feel are valuable experimental reactions and quantities.\cite{EXFOR,EXFORHistory} 

The basic unit of EXFOR is an {\tt ENTRY}. An {\tt ENTRY} corresponds to one experiment and contains the numerical data, the related bibliographic information and a brief description of the experimental method. What an EXFOR {\tt ENTRY} really represents is the results of work that was performed at a given laboratory at a given time. An {\tt ENTRY} does not necessarily correspond to one particular publication, but very often corresponds to several publications. The EXFOR compiler takes the essential information from all available sources, and if needed, contacts the author for more information.

An entry is typically divided in several {\tt SUBENT}s containing the data tables resulting from the experiment.  Each {\tt SUBENT} contains a {\tt REACTION} field which encodes what reaction was studied (e.g. $^1$H($n$,el)) and what quantity was measured (e.g. cross-section or angular distribution).  A {\tt SUBENT} may also contain a {\tt MONITOR} field which encodes one or more well characterized reactions and quantities used to reduce or eliminate systematic experimental errors.  Thus, reaction monitors are an important part of experimental data reduction.  Often the measured data encoded in the {\tt REACTION} field is measured relative to the reaction/quantity encoded in the {\tt MONITOR} field.  There is usually a straightforward mapping between the reactions/quantities measured in EXFOR and the evaluated reactions/quantities stored in the ENDF libraries.

Several specific reaction/quantities are important enough, usually because of one or more specific applications, that the nuclear data community has elevated them to the level of an international reference standard.  Many experimenters use these reaction/quantities as monitors in their experiments. References \cite{Standards,Atlas,IAEAMedical,IRDFF}  provide details of well known neutron-induced, charged-particle and photonuclear standard reaction/quantities. We divided these references into two different classes. We define Tier 1 observables as the product of sustained evaluation efforts, with periodic refinement.  Our Tier 1 standards include the evaluations from the ENDF/B Neutron Standards \cite{Standards} project and the {\em Atlas of Neutron Resonances} \cite{Atlas}.  Our second tier encompasses standards that are of very high quality but are not performed as part of a sustained project.  There may be follow ups or limited refinements.  This second tier includes medical and dosimitry evaluations in Ref. \cite{IAEAMedical} and the results of the IRDFF project \cite{IRDFF}.  There is also a new Tier 1 standards-level effort just beginning known as CIELO pilot project \cite{CIELO}.   CIELO promises to generate entire standards-level evaluations including all reactions/quantities needed for the ENDF-formatted libraries for neutron-induced reactions on $^{1}$H, $^{16}$O, $^{56}$Fe, $^{235}$U, $^{238}$U and $^{239}$Pu. 

When undertaking this project we specifically set out with the goal of answering some important questions. 
\begin{itemize}
\item What are the most connected targets? What are the most connected reactions/quantities?
\item Are there reactions/quantities that are so connected that they should be a standard?
\item What is the connection number distribution for targets and reactions?  What is the link number distribution between any two targets?
\item Can we use this information to inform new measurements that decrease the distance between important targets and standards?
\item Are there ``bottlenecks'' along the pathways from a given reaction to reaction standards that are not well measured?
\item What elements and isotopes of reactions are not linked? Are any of them important for specific applications?
\end{itemize}
In order to attempt to resolve these questions we utilized graph theory as a tool to examine the connections between measurements in EXFOR. 

In this work, we take an abstract view of the EXFOR database and generate an undirected graph describing all the connections between reactions/quantities in the EXFOR database.  From just these connections, we can infer what reactions/quantities the nuclear science community collectively (and somewhat unconsciously) views as important.  This set of reactions/quantities does not exactly match our previous expectations.  We will provide a series of recommendations for reactions/quantities that should be considered for elevation to the level of the standards in references \cite{Standards,Atlas,IAEAMedical,IRDFF} or possibly included in a follow-on CIELO project.  We also find that our graph is disconnected, implying there are large numbers of reactions/quantities that are not pinned to any monitor.  In many cases, this is due to poor coding of the  EXFOR {\tt ENTRY}s.  Although this is a serious problem in our study, there is no easy fix.   Even if additional information is given, it is often given in one of the free text fields in EXFOR which are difficult, if not impossible, to parse.

\section{Constructing the graph}
We used the {\tt x4i} code \cite{x4i} to read the EXFOR database and parse the {\tt REACTION} and {\tt MONITOR} strings.  We then built up the undirected graph within {\tt x4i} and stored the resulting graph in a GraphML formatted file.  In this section we detail how we construct the graph.

\begin{table}[ht]
\centering
\begin{tabular}{lc}
\hline\hline
Description		& Examples\\ 
\hline\hline
regular node		& \parbox[c]{2.5cm}{\includegraphics[width=0.5cm]{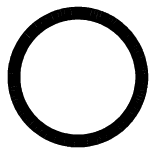}}\\
isomer target	& \parbox[c]{2.5cm}{\includegraphics[width=0.5cm]{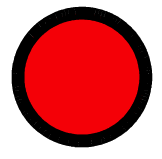}}\\
elemental target	& \parbox[c]{2.5cm}{\includegraphics[width=0.5cm]{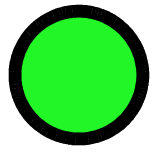}}\\ \hline
ENDF/B-VII.1 Neutron Standard \cite{Standards}& \parbox[c]{2.5cm}{\includegraphics[width=0.5cm]{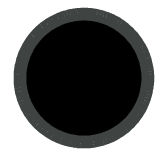}}\\
{\em Atlas of Neutron Resonances} standard \cite{Atlas}& \parbox[c]{2.5cm}{\includegraphics[width=0.5cm]{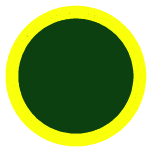}}\\ \hline
IAEA Medical/Dosimiter \cite{IAEAMedical}& \parbox[c]{2.5cm}{\includegraphics[width=0.5cm]{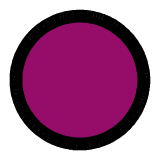}}\\
IRDFF \cite{IRDFF}& \parbox[c]{2.5cm}{\includegraphics[width=0.5cm]{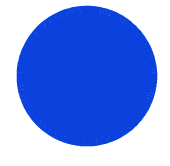}}\\\hline
CIELO target \cite{CIELO}&\parbox[c]{2.5cm}{\includegraphics[width=0.5cm]{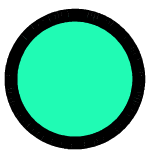}}\\ \hline
Proposed standard (ours or from Ref. \cite{IAEAStdsMeeting2013}) & \parbox[c]{2.5cm}{\includegraphics[width=0.5cm]{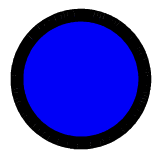}}\\ \hline
\end{tabular}
\caption{\label{table:nodes}Types of nodes in our graph.}
\end{table}

\subsection{Types of nodes/vertices}
Each EXFOR {\tt SUBENT} corresponds to one or more measured datasets and each dataset in the {\tt SUBENT} is associated with exactly one reaction/quantity in the {\tt REACTION} field.  Multiple reaction/quantities and datasets are denoted with EXFOR pointers.   Each {\tt SUBENT} may also contain a {\tt MONITOR} field which we also note.  Both {\tt REACTION} and {\tt MONITOR} fields have essentially the same format and contain much the same information \cite{EXFORManual}.  The {\tt MONITOR} field may  also contain other free-text information detailing how the monitor was used and we ignore this information.  An example of a simple measurement is 
\begin{verbatim}
    (1-H-1(N,TOT),,SIG,,MXW)
\end{verbatim}
This {\tt REACTION} field tells us that the $^1$H($n$,tot) Maxwellian ({\tt MXW}) averaged cross-section ({\tt SIG}) was measured in the associated {\tt SUBENT}.  In our graph, we assign each of these elementary reaction/quantities in {\tt REACTION} or {\tt MONITOR} fields to a node.   

While building our graph, we associate the number of occurrences of each elementary reaction/quantity with the corresponding node. In our graph we did not make any distinction between variations in observerables.  For example, ``PAR,SIG'' coding for partial cross sections, ``SIG'' for integrated cross sections and ``CN,SIG'' for compound nuclear cross sections are all treated as ``SIG''. 

We make two notes on double counting of nodes.  First, when the EXFOR compiler flags a reaction/quantity in the {\tt MONITOR} field and also compiles the reaction/quantity as a ratio we count both the occurrences of the monitor reaction/quantity separately.  This is straightforward to fix and will be done in future incarnations of this project.   Second, it often happens that an experimenter who makes a ratio measurement will publish both the ratio values and the unfolded absolute value of a measured reaction/quantity.  An EXFOR compiler will then compile both values as if they are independent datasets and provide an explanation of the sets in one of the EXFOR free text fields.  As one must parse the free text descriptions of the experiment in order to discern this, we have no simple workaround.  

Table \ref{table:nodes} lists all of the types of nodes.  In this table, the nodes are colorized by whether they correspond to one of the standards in the various standards efforts \cite{Standards,Atlas,IAEAMedical,IRDFF}.  The specific reactions/quantities included from these various standards projects are given in Tables \ref{table:tier1Standards}--\ref{table:proposedStandards}. 

\begin{table*}[ht]
\caption{\label{table:tier1Standards}Table of all Tier 1 standards.  References marked with a ``*'' indicate that the reaction/quantity are a by-product of the evaluation process.}
\squeezetable
\begin{tabular}{lll}
\hline\hline
Reaction & Observable & Reference\\
\hline\hline
$^{1}$H($n$, el) & $\sigma$ & ENDF/B-VII.1 Standard \cite{Standards}, CIELO \cite{CIELO}\\
$^{3}$He($n$, $p$) & $\sigma$ & ENDF/B-VII.1 Standard \cite{Standards}\\
$^{6}$Li($n$, $t$) & $\sigma$ & ENDF/B-VII.1 Standard \cite{Standards}, IRDFF \cite{IRDFF}\\
$^{10}$B($n$, $\alpha$+$\gamma$) & $\sigma$ & ENDF/B-VII.1 Standard \cite{Standards}\\
$^{nat}$C($n$, el) & $\sigma$ & ENDF/B-VII.1 Standard \cite{Standards}\\
$^{197}$Au($n$, $\gamma$) & $\sigma$ & ENDF/B-VII.1 Standard \cite{Standards}, Atlas \cite{Atlas}, IRDFF \cite{IRDFF}\\
$^{235}$U($n$, f) & $\sigma$ & ENDF/B-VII.1 Standard \cite{Standards}, CIELO \cite{CIELO}\\
$^{10}$B($n$, $\alpha$) & $\sigma$ & ENDF/B-VII.1 Standard \cite{Standards}, IRDFF \cite{IRDFF}\\
$^{238}$U($n$, f) & $\sigma$ & ENDF/B-VII.1 Standard \cite{Standards}, CIELO \cite{CIELO}\\
$^{10}$B($n$, $\alpha$+$\gamma$) & $\sigma$ & ENDF/B-VII.1 Standard \cite{Standards}\\
$^{6}$Li($n$, el) & $\sigma$ & ENDF/B-VII.1 Standard$^*$ \cite{Standards}\\
$^{10}$B($n$, el) & $\sigma$ & ENDF/B-VII.1 Standard$^*$ \cite{Standards}\\
$^{239}$Pu($n$, f) & $\sigma$ & ENDF/B-VII.1 Standard$^*$ \cite{Standards}, CIELO \cite{CIELO}\\
$^{238}$U($n$, $\gamma$) & $\sigma$ & ENDF/B-VII.1 Standard$^*$ \cite{Standards}, IRDFF \cite{IRDFF}, CIELO \cite{CIELO}\\
$^{10}$B($n$, $\alpha$) & $\sigma$ & ENDF/B-VII.1 Standard$^*$ \cite{Standards}\\
$^{197}$Au($n$, $\gamma$) & RI & Atlas \cite{Atlas}\\
$^{197}$Au($n$, $\gamma$) & $\sigma$ & ENDF/B-VII.1 Standard \cite{Standards}, Atlas \cite{Atlas}, IRDFF \cite{IRDFF}\\
$^{59}$Co($n$, $\gamma$) & $\sigma$ & Atlas \cite{Atlas}, IRDFF \cite{IRDFF}\\
$^{59}$Co($n$, $\gamma$) & RI & Atlas \cite{Atlas}\\
$^{55}$Mn($n$, $\gamma$) & $\sigma$ & Atlas \cite{Atlas}, IRDFF \cite{IRDFF}\\
$^{55}$Mn($n$, $\gamma$) & RI & Atlas \cite{Atlas}\\
$^{35}$Cl($n$, $\gamma$) & $\sigma$ & Atlas \cite{Atlas}\\
$^{10}$B($n$, $\gamma$) & $\sigma$ & Atlas \cite{Atlas}\\
$^{nat}$B($n$, abs) & $\sigma$ & Atlas \cite{Atlas}\\
$^{6}$Li($n$, $\alpha$) & $\sigma$ & Atlas \cite{Atlas}\\
$^{nat}$Li($n$, abs) & $\sigma$ & Atlas \cite{Atlas}\\
$^{1}$H($n$, $\gamma$) & $\sigma$ & Atlas \cite{Atlas}, CIELO \cite{CIELO}\\
\hline
\end{tabular}
\end{table*}

{\bf Tier 1 projects} are detailed here and summarized in Table \ref{table:tier1Standards}:
\begin{itemize}
\item The Neutron Standards Project \cite{Standards} is an ongoing project under the mandates of a long term IAEA Coordinated Research Project (CRP) and the Cross Section Evaluation Working Group (CSEWG).  The Neutron Standards project has been operating since the first ENDF library, ENDF/B-I in 1968. The project's long term goal is to deliver international standards level evaluated cross section tables for specific reactions (see Table  \ref{table:tier1Standards}).  Over the years, other reaction/quantities have been proposed for addition to the Neutron Standards Project.  One set seems poised to become port of the international standards (see Table \ref{table:proposedStandards}) as was discussed at the July 2013 IAEA Technical Meeting \cite{IAEAStdsMeeting2013}.
\item The {\em Atlas of Neutron Resonances} \cite{Atlas} is an encyclopedic compilation of nuclear cross section resonance parameters from S. Mughabghab.  This book has been continuously updated since 1952 and is now in its $5^{th}$ edition.  Page 8 of the latest edition includes a table of standards, summarized in Table \ref{table:tier1Standards}.  These standards are values for resonance integrals and thermal cross sections.
\end{itemize}

\begin{table}[ht]
\caption{\label{table:proposedStandards}Table of possible standards proposed in the July 2013 IAEA Technical Meeting (see Ref. \cite{IAEAStdsMeeting2013}).}
\begin{tabular}{lll}
\hline\hline
Reaction & Observable & Reference\\
\hline\hline
$^{252}$Cf($0$, f) & $\bar{\nu}$ & Proposed \cite{IAEAStdsMeeting2013}\\
$^{252}$Cf($0$, f) & $d\nu/dE'$ & Proposed \cite{IAEAStdsMeeting2013}\\
$^{27}$Al($n$, $\alpha$) & $\sigma$ & Proposed \cite{IAEAStdsMeeting2013}, IRDFF \cite{IRDFF}\\
$^{209}$Bi($n$, f) & $\sigma$ & Proposed \cite{IAEAStdsMeeting2013}\\
\hline
\end{tabular}
\end{table}

\begin{table}[ht]
\caption{\label{table:tier2aStandards}Table of Medical/Dosimeter \cite{IAEAMedical} Tier 2 standards.}
\squeezetable
\begin{tabular}{lll}
\hline\hline
Reaction & Observable & Reference\\
\hline\hline
$^{27}$Al($p$, X+$^{22}$Na) & $\sigma$ & Medical/Dosimeter \cite{IAEAMedical}\\
$^{27}$Al($p$, X+$^{24}$Na) & $\sigma$ & Medical/Dosimeter \cite{IAEAMedical}\\
$^{nat}$Ti($p$, X+$^{48}$V) & $\sigma$ & Medical/Dosimeter \cite{IAEAMedical}\\
$^{nat}$Ni($p$, X+$^{57}$Ni) & $\sigma$ & Medical/Dosimeter \cite{IAEAMedical}\\
$^{nat}$Cu($p$, X+$^{56}$Co) & $\sigma$ & Medical/Dosimeter \cite{IAEAMedical}\\
$^{nat}$Cu($p$, X+$^{62}$Zn) & $\sigma$ & Medical/Dosimeter \cite{IAEAMedical}\\
$^{nat}$Cu($p$, X+$^{63}$Zn) & $\sigma$ & Medical/Dosimeter \cite{IAEAMedical}\\
$^{nat}$Cu($p$, X+$^{65}$Zn) & $\sigma$ & Medical/Dosimeter \cite{IAEAMedical}\\
$^{27}$Al($d$, X+$^{22}$Na) & $\sigma$ & Medical/Dosimeter \cite{IAEAMedical}\\
$^{27}$Al($d$, X+$^{24}$Na) & $\sigma$ & Medical/Dosimeter \cite{IAEAMedical}\\
$^{nat}$Ti($d$, X+$^{48}$V) & $\sigma$ & Medical/Dosimeter \cite{IAEAMedical}\\
$^{nat}$Fe($d$, X+$^{56}$Co) & $\sigma$ & Medical/Dosimeter \cite{IAEAMedical}\\
$^{nat}$Ni($d$, X+$^{61}$Cu) & $\sigma$ & Medical/Dosimeter \cite{IAEAMedical}\\
$^{27}$Al($^{3}$He, X+$^{22}$Na) & $\sigma$ & Medical/Dosimeter \cite{IAEAMedical}\\
$^{27}$Al($^{3}$He, X+$^{24}$Na) & $\sigma$ & Medical/Dosimeter \cite{IAEAMedical}\\
$^{nat}$Ti($^{3}$He, X+$^{48}$V) & $\sigma$ & Medical/Dosimeter \cite{IAEAMedical}\\
$^{27}$Al($\alpha$, X+$^{22}$Na) & $\sigma$ & Medical/Dosimeter \cite{IAEAMedical}\\
$^{27}$Al($\alpha$, X+$^{24}$Na) & $\sigma$ & Medical/Dosimeter \cite{IAEAMedical}\\
$^{nat}$Ti($\alpha$, X+$^{48}$V) & $\sigma$ & Medical/Dosimeter \cite{IAEAMedical}\\
$^{nat}$Cu($\alpha$, X+$^{66}$Ga) & $\sigma$ & Medical/Dosimeter \cite{IAEAMedical}\\
$^{nat}$Cu($\alpha$, X+$^{67}$Ga) & $\sigma$ & Medical/Dosimeter \cite{IAEAMedical}\\
$^{nat}$Cu($\alpha$, X+$^{65}$Zn) & $\sigma$ & Medical/Dosimeter \cite{IAEAMedical}\\
$^{15}$N($p$, $n$+$^{15}$0) & $\sigma$ & Medical/Dosimeter \cite{IAEAMedical}\\
$^{18}$O($p$, $n$+$^{18}$F) & $\sigma$ & Medical/Dosimeter \cite{IAEAMedical}\\
$^{67}$Zn($p$, $n$+$^{67}$Ga) & $\sigma$ & Medical/Dosimeter \cite{IAEAMedical}\\
$^{68}$Zn($p$, $2n$+$^{67}$Ga) & $\sigma$ & Medical/Dosimeter \cite{IAEAMedical}\\
$^{111}$Cd($p$, $n$+$^{111}$In) & $\sigma$ & Medical/Dosimeter \cite{IAEAMedical}\\
$^{112}$Cd($p$, $2n$+$^{111}$In) & $\sigma$ & Medical/Dosimeter \cite{IAEAMedical}\\
$^{123}$Te($p$, $n$+$^{123}$I) & $\sigma$ & Medical/Dosimeter \cite{IAEAMedical}\\
$^{124}$Te($p$, $2n$+$^{123}$I) & $\sigma$ & Medical/Dosimeter \cite{IAEAMedical}\\
$^{124}$Te($p$, $n$+$^{124}$I) & $\sigma$ & Medical/Dosimeter \cite{IAEAMedical}\\
$^{127}$I($p$, $5n$+$^{123}$Xe) & $\sigma$ & Medical/Dosimeter \cite{IAEAMedical}\\
$^{127}$I($p$, $3n$+$^{125}$Xe) & $\sigma$ & Medical/Dosimeter \cite{IAEAMedical}\\
$^{nat}$Kr($p$, X+$^{81}$Rb) & $\sigma$ & Medical/Dosimeter \cite{IAEAMedical}\\
$^{82}$K4($p$, $2n$+$^{82}$Rb) & $\sigma$ & Medical/Dosimeter \cite{IAEAMedical}\\
$^{124}$Xe($p$, $2n$+$^{123}$Cs) & $\sigma$ & Medical/Dosimeter \cite{IAEAMedical}\\
$^{124}$Xe($p$, $p$+$n$+$^{123}$Xe) & $\sigma$ & Medical/Dosimeter \cite{IAEAMedical}\\
$^{203}$Tl($p$, $2n$+$^{201}$Pb) & $\sigma$ & Medical/Dosimeter \cite{IAEAMedical}\\
$^{203}$Tl($p$, $2n$+$^{202m}$Pb) & $\sigma$ & Medical/Dosimeter \cite{IAEAMedical}\\
$^{203}$Tl($p$, $4n$+$^{200}$Pb) & $\sigma$ & Medical/Dosimeter \cite{IAEAMedical}\\
$^{14}$N($p$, $\alpha$+$^{11}$C) & $\sigma$ & Medical/Dosimeter \cite{IAEAMedical}\\
$^{69}$Ga($p$, $2n$+$^{68}$Ge) & $\sigma$ & Medical/Dosimeter \cite{IAEAMedical}\\
$^{nat}$Ga($p$, X+$^{68}$Ge) & $\sigma$ & Medical/Dosimeter \cite{IAEAMedical}\\
$^{16}$O($p$, $\alpha$+$^{13}$N) & $\sigma$ & Medical/Dosimeter \cite{IAEAMedical}\\
$^{85}$Rb($p$, $4n$+$^{82}$Sr) & $\sigma$ & Medical/Dosimeter \cite{IAEAMedical}\\
$^{nat}$Pb($p$, X+$^{82}$Sr) & $\sigma$ & Medical/Dosimeter \cite{IAEAMedical}\\
$^{14}$N($d$, $n$+$^{15}$0) & $\sigma$ & Medical/Dosimeter \cite{IAEAMedical}\\
$^{nat}$Ne($d$, X+$^{19}$F) & $\sigma$ & Medical/Dosimeter \cite{IAEAMedical}\\
\hline
\end{tabular}
\end{table}

{\bf Tier 2 projects} are
\begin{itemize}
\item In 2003 an IAEA CRP titled ``Charged-particle cross section database for medical radioisotope production; Diagnostic radioisotopes and monitor reactions'' \cite{IAEAMedical} concluded.  The aim of the project was to compile evaluated data for both radioisotope production and for use as reaction monitors.  The list of reactions/quantities covered is given in Table \ref{table:tier2aStandards}.  The project began in 1995 and covered three major areas:
	\begin{itemize}
	\item Monitors for light ion charged particle beams
	\item Cross section data on the most commonly used radioisotopes in medicine
	\item Uncertainty on the tabulated data
	\end{itemize} 
A large amount of the data added to the CRP is in the 30 MeV range or higher.   
\item The International Reactor Dosimetry and Fusion File (IRDFF) \cite{IRDFF} is a standards level set of evaluated cross sections and fission product yields (FPY) for use as neutron dosimetry reactions.  All 66 cross section and FPY sets in IRDFF contain covariance data.  The 2012 version of IRDFF is the second release of the IRDFF.  The data in IRDFF is summarized in Table \ref{table:tier2bStandards}.   
\end{itemize}
It should be noted that the Ion Beam Analysis Nuclear Data Library (IBANDL) \cite{IBANDL} is a library of nuclear cross section data relevant to ion beam analysis.  IBANDL is not included as a standard here since it contains only experimental data and is already included in EXFOR. 

In addition to these Tier 1 and 2 standards, there is a new project, the Collaborative International Evaluated Library Organization (CIELO) project with the goal of producing entire standards level evaluations for $^1$H, $^{16}$O, $^{56}$Fe, $^{235,238}$U and $^{239}$Pu \cite{CIELO}.  An entire evaluation is one in which all cross sections and all outgoing particle distributions on all open channels with threshold between $10^{-5}$ eV and 20 MeV are given.   The CIELO project is just beginning but should yield first results by 2016.

\begin{longtable}{lp{1.7cm}p{3.5cm}}
\caption{\label{table:tier2bStandards}Table of IRDFF \cite{IRDFF} Tier 2 standards.}\\
\hline\hline
Reaction & Observable & Reference\\
\hline\hline
\endhead
\endfoot
$^{6}$Li($n$, $t$) & $\sigma$ & ENDF/B-VII.1 Standard \cite{Standards}, IRDFF \cite{IRDFF}\\
$^{10}$B($n$, $\alpha$) & $\sigma$ & ENDF/B-VII.1 Standard \cite{Standards}, IRDFF \cite{IRDFF}\\
$^{19}$F($n$, $2n$) & $\sigma$ & IRDFF \cite{IRDFF}\\
$^{23}$Na($n$, $2n$) & $\sigma$ & IRDFF \cite{IRDFF}\\
$^{23}$Na($n$, $\gamma$) & $\sigma$ & IRDFF \cite{IRDFF}\\
$^{24}$Mg($n$, $p$) & $\sigma$ & IRDFF \cite{IRDFF}\\
$^{27}$Al($n$, $p$) & $\sigma$ & IRDFF \cite{IRDFF}\\
$^{27}$Al($n$, $\alpha$) & $\sigma$ & Proposed \cite{IAEAStdsMeeting2013}, IRDFF \cite{IRDFF}\\
$^{31}$P($n$, $p$) & $\sigma$ & IRDFF \cite{IRDFF}\\
$^{32}$S($n$, $p$) & $\sigma$ & IRDFF \cite{IRDFF}\\
$^{45}$Sc($n$, $\gamma$) & $\sigma$ & IRDFF \cite{IRDFF}\\
$^{46}$Ti($n$, $2n$) & $\sigma$ & IRDFF \cite{IRDFF}\\
$^{46}$Ti($n$, $p$) & $\sigma$ & IRDFF \cite{IRDFF}\\
$^{47}$Ti($n$, X) & $\sigma$ & IRDFF \cite{IRDFF}\\
$^{47}$Ti($n$, $p$) & $\sigma$ & IRDFF \cite{IRDFF}\\
$^{48}$Ti($n$, X) & $\sigma$ & IRDFF \cite{IRDFF}\\
$^{48}$Ti($n$, $p$) & $\sigma$ & IRDFF \cite{IRDFF}\\
$^{49}$Ti($n$, X) & $\sigma$ & IRDFF \cite{IRDFF}\\
$^{51}$V($n$, $\alpha$) & $\sigma$ & IRDFF \cite{IRDFF}\\
$^{52}$Cr($n$, $2n$) & $\sigma$ & IRDFF \cite{IRDFF}\\
$^{55}$Mn($n$, $\gamma$) & $\sigma$ & Atlas \cite{Atlas}, IRDFF \cite{IRDFF}\\
$^{55}$Mn($n$, $2n$) & $\sigma$ & IRDFF \cite{IRDFF}\\
$^{54}$Fe($n$, $2n$) & $\sigma$ & IRDFF \cite{IRDFF}\\
$^{54}$Fe($n$, $p$) & $\sigma$ & IRDFF \cite{IRDFF}\\
$^{54}$Fe($n$, $\alpha$) & $\sigma$ & IRDFF \cite{IRDFF}\\
$^{56}$Fe($n$, $p$) & $\sigma$ & IRDFF \cite{IRDFF}, CIELO \cite{CIELO}\\
$^{58}$Fe($n$, $\gamma$) & $\sigma$ & IRDFF \cite{IRDFF}\\
$^{59}$Co($n$, $2n$) & $\sigma$ & IRDFF \cite{IRDFF}\\
$^{59}$Co($n$, $3n$) & $\sigma$ & IRDFF \cite{IRDFF}\\
$^{59}$Co($n$, $\gamma$) & $\sigma$ & Atlas \cite{Atlas}, IRDFF \cite{IRDFF}\\
$^{59}$Co($n$, $p$) & $\sigma$ & IRDFF \cite{IRDFF}\\
$^{59}$Co($n$, $\alpha$) & $\sigma$ & IRDFF \cite{IRDFF}\\
$^{58}$Ni($n$, $2n$) & $\sigma$ & IRDFF \cite{IRDFF}\\
$^{58}$Ni($n$, $p$) & $\sigma$ & IRDFF \cite{IRDFF}\\
$^{60}$Ni($n$, $p$) & $\sigma$ & IRDFF \cite{IRDFF}\\
$^{63}$Cu($n$, $2n$) & $\sigma$ & IRDFF \cite{IRDFF}\\
$^{63}$Cu($n$, $\gamma$) & $\sigma$ & IRDFF \cite{IRDFF}\\
$^{63}$Cu($n$, $\alpha$) & $\sigma$ & IRDFF \cite{IRDFF}\\
$^{65}$Cu($n$, $2n$) & $\sigma$ & IRDFF \cite{IRDFF}\\
$^{64}$Zn($n$, $p$) & $\sigma$ & IRDFF \cite{IRDFF}\\
$^{67}$Za($n$, $p$) & $\sigma$ & IRDFF \cite{IRDFF}\\
$^{75}$As($n$, $2n$) & $\sigma$ & IRDFF \cite{IRDFF}\\
$^{89}$Y($n$, $2n$) & $\sigma$ & IRDFF \cite{IRDFF}\\
$^{90}$Zr($n$, $2n$) & $\sigma$ & IRDFF \cite{IRDFF}\\
$^{92}$Mo($n$, $p$) & $\sigma$ & IRDFF \cite{IRDFF}\\
$^{93}$Nb($n$, $2n$) & $\sigma$ & IRDFF \cite{IRDFF}\\
$^{93}$Nb($n$, $2n$+$^{92m}$Nb) & $\sigma$ & IRDFF \cite{IRDFF}\\
$^{93}$Nb($n$, inel) & $\sigma$ & IRDFF \cite{IRDFF}\\
$^{93}$Nb($n$, $\gamma$) & $\sigma$ & IRDFF \cite{IRDFF}\\
$^{103}$Rh($n$, inel+$^{103m}$Rh) & $\sigma$ & IRDFF \cite{IRDFF}\\
$^{109}$Ag($n$, $\gamma$+$^{110m}$Ag) & $\sigma$ & IRDFF \cite{IRDFF}\\
$^{113}$In($n$, inel) & $\sigma$ & IRDFF \cite{IRDFF}\\
$^{113}$In($n$, inel+$^{113m}$In) & $\sigma$ & IRDFF \cite{IRDFF}\\
$^{115}$In($n$, $2n$+$^{114m}$In) & $\sigma$ & IRDFF \cite{IRDFF}\\
$^{115}$In($n$, inel) & $\sigma$ & IRDFF \cite{IRDFF}\\
$^{115}$In($n$, inel+$^{115m}$In) & $\sigma$ & IRDFF \cite{IRDFF}\\
$^{115}$In($n$, $\gamma$+$^{116m}$In) & $\sigma$ & IRDFF \cite{IRDFF}\\
$^{127}$I($n$, $2n$) & $\sigma$ & IRDFF \cite{IRDFF}\\
$^{139}$La($n$, $\gamma$) & $\sigma$ & IRDFF \cite{IRDFF}\\
$^{141}$Pr($n$, $2n$) & $\sigma$ & IRDFF \cite{IRDFF}\\
$^{169}$Tm($n$, $2n$) & $\sigma$ & IRDFF \cite{IRDFF}\\
$^{169}$Tm($n$, $3n$) & $\sigma$ & IRDFF \cite{IRDFF}\\
$^{181}$Ta($n$, $\gamma$) & $\sigma$ & IRDFF \cite{IRDFF}\\
$^{186}$W($n$, $\gamma$) & $\sigma$ & IRDFF \cite{IRDFF}\\
$^{197}$Au($n$, $2n$) & $\sigma$ & IRDFF \cite{IRDFF}\\
$^{197}$Au($n$, $\gamma$) & $\sigma$ & ENDF/B-VII.1 Standard \cite{Standards}, Atlas \cite{Atlas}, IRDFF \cite{IRDFF}\\
$^{199}$Hg($n$, inel+$^{199m}$Hg) & $\sigma$ & IRDFF \cite{IRDFF}\\
$^{204}$Pb($n$, inel+$^{204m}$Pb) & $\sigma$ & IRDFF \cite{IRDFF}\\
$^{209}$Bi($n$, $3n$) & $\sigma$ & IRDFF \cite{IRDFF}\\
$^{232}$Th($n$, f) & FPY (ELEM/MASS) & IRDFF \cite{IRDFF}\\
$^{232}$Th($n$, $\gamma$) & $\sigma$ & IRDFF \cite{IRDFF}\\
$^{235}$U($n$, f) & FPY (ELEM/MASS) & IRDFF \cite{IRDFF}\\
$^{238}$U($n$, f) & FPY (ELEM/MASS) & IRDFF \cite{IRDFF}\\
$^{238}$U($n$, $\gamma$) & $\sigma$ & ENDF/B-VII.1 Standard$^*$ \cite{Standards}, IRDFF \cite{IRDFF}, CIELO \cite{CIELO}\\
$^{237}$Np($n$, f) & FPY (ELEM/MASS) & IRDFF \cite{IRDFF}\\
$^{239}$Pu($n$, f) & FPY (ELEM/MASS) & IRDFF \cite{IRDFF}\\
$^{241}$Am($n$, f) & FPY (ELEM/MASS) & IRDFF \cite{IRDFF}\\
\hline
\end{longtable}

\subsection{Types of connections/edges}

The nodes in our graph are connected by edges.  The types of edges we consider are listed in Table \ref{table:edges}.  By far the most common type of edge in our graph is the {\tt MONITOR}--{\tt REACTION} connection.  However, the EXFOR format provides several other connections between elementary nodes.  {\tt REACTION} and {\tt MONITOR} fields may also contain mathematical relations, e.g.
\begin{verbatim}
    ((94-PU-239(N,F),,NU,,MXW)/
        (92-U-235(N,F),,NU,,MXW))
\end{verbatim}
In this example, the measurement was the ratio of Maxwellian averaged $\overline{\nu}_p$'s from $^{239}$Pu($n$,f) and $^{235}$U($n$,f).
Any relation using {\tt +},{\tt -},{\tt *},{\tt /}, {\tt //}, and {\tt =} are allowed in the {\tt REACTION} and {\tt MONITOR} fields (here {\tt //} means double ratio).  EXFOR also allows what we call ``isomer math'':
\begin{verbatim}
    (72-HF-177(N,G)72-HF-178-M/T,,SIG/RAT)
\end{verbatim}
Here, what was measured was the ratio of $^{177}$Hf($n,\gamma$)$^{178m}$Hf cross section to the total of $^{177}$Hf($n,\gamma$)$^{178m}$Hf and $^{177}$Hf($n,\gamma$)$^{178g}$Hf cross sections. 
 
We link all of the Neutron Standards \cite{Standards} reactions/quantities together because they are evaluated simultaneously.  We also consider all reactions/quantities covered by one isotope in the CIELO pilot project \cite{CIELO}  to be linked together since they to are evaluated together.  

\begin{table*}[ht]
\caption{\label{table:edges}Types of edges in our graph.}
\centering
\begin{tabular}{p{2in}p{3in}p{3.5cm}}
\hline\hline
Type\ 	& Description & Example\\ 
\hline\hline
Mathematical relations &
	These types of connections can be a simple ratio or a more complex mathematical relations between two or more other nodes.  These include ``isomer math'' and the special quantities and sum rules in Table \ref{table:motifs}.  &
	\parbox[c]{3.5cm}{\includegraphics[width=2.5cm]{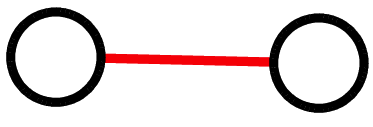}}\\
\hline 
Monitor			& 
Typically a second, well characterized target used to reduce or eliminate systematic experimental problems during data 	analysis.	& 
	\parbox[c]{3.5cm}{\includegraphics[width=2.5cm]{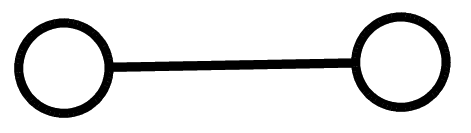}}\\
\hline
Elemental		& 
	Data from a natural element is connected to every stable isotope of the element for the same measurement. & 	
	\parbox[c]{3.5cm}{\includegraphics[width=2.5cm]{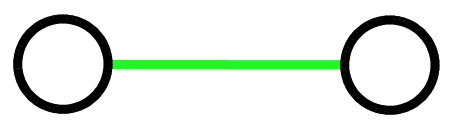}}\\  
\hline
Neutron 	Standards/CIELO		& 
	All reactions/quantities are evaluated simultaneously and therefore are linked.& 
    \parbox[c]{3.5cm}{\includegraphics[width=2.5cm]{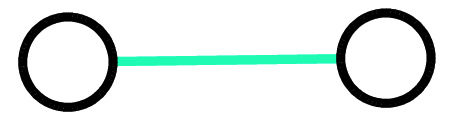}}\\
\hline
\end{tabular}
\end{table*}

\subsection{Other connections and graph motifs}
Many graphs have motifs or common or repeated patterns. Our network also has motifs: we impose several on the graph as a result of our coding for some of the more obscure EXFOR reaction/quantities.  Table \ref{table:motifs} shows the clusters of links for EXFOR quantities {\tt ALF}, {\tt ETA} and {\tt RI} and EXFOR reactions {\tt NON}, {\tt INEL} and {\tt SCT}.  Finally, as an element is an abundance weighted sum of the isotopes that make up the element, we link any reaction/quantity on an elemental target to the corresponding isotopic reaction/quantities.  The motifs are described and depicted in Table \ref{table:motifs}.

\begin{table*}[ht]
\caption{\label{table:motifs}Motifs in the graph, these also happen to be connections we wanted to capture in the graph.} 
\centering
\begin{tabular}{cccc}
\hline\hline
EXFOR Quantity\ 	& Definition\ 					& Example\\[0.5ex]
\hline\hline
ALF 				& $\alpha \equiv  \sigma_{\gamma}/\sigma_{f}$ 	& \parbox[c]{6.5cm}{\includegraphics[width=6cm]{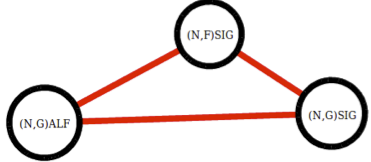}} \\ \hline 
ETA 				& $\eta \equiv  \overline{\nu}\sigma_{f}/
					(\sigma_{\gamma}+\sigma_{f})$	& \parbox[c]{5.5cm}{\includegraphics[width=5cm]{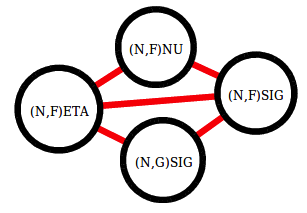}} \\ \hline
SCT 				& $\sigma_{sct} \equiv \sigma_{el}+\sigma_{inel}$ 	& \parbox[c]{5.5cm}{\includegraphics[width=5cm]{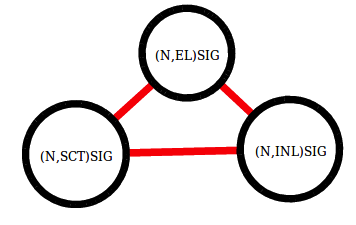}} \\ \hline
NON 				& $\sigma_{non} \equiv  \sigma_{tot}-\sigma_{el}$ 		& \parbox[c]{5.5cm}{\includegraphics[width=5cm]{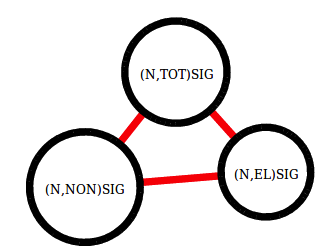}} \\ \hline
RI 				& $RI \equiv  \displaystyle\int_0^\infty dE			
					\frac{\sigma(E)}{E}$   		& \parbox[c]{6.5cm}{\includegraphics[width=6cm]{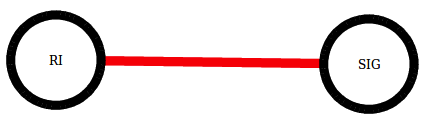}}\\ \hline
Elemental 		& $\sigma^{nat} \equiv \sum_i w_i \sigma^i(E)$		& \parbox[c]{5.5cm}{\includegraphics[width=5cm]{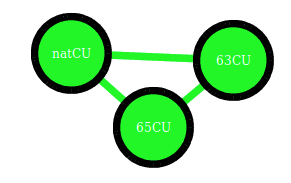}} \\ 
[1ex]
\hline
\end{tabular}
\end{table*}

%

\section{Gross features of our graph}
\begin{figure*}
\includegraphics[width=\textwidth]{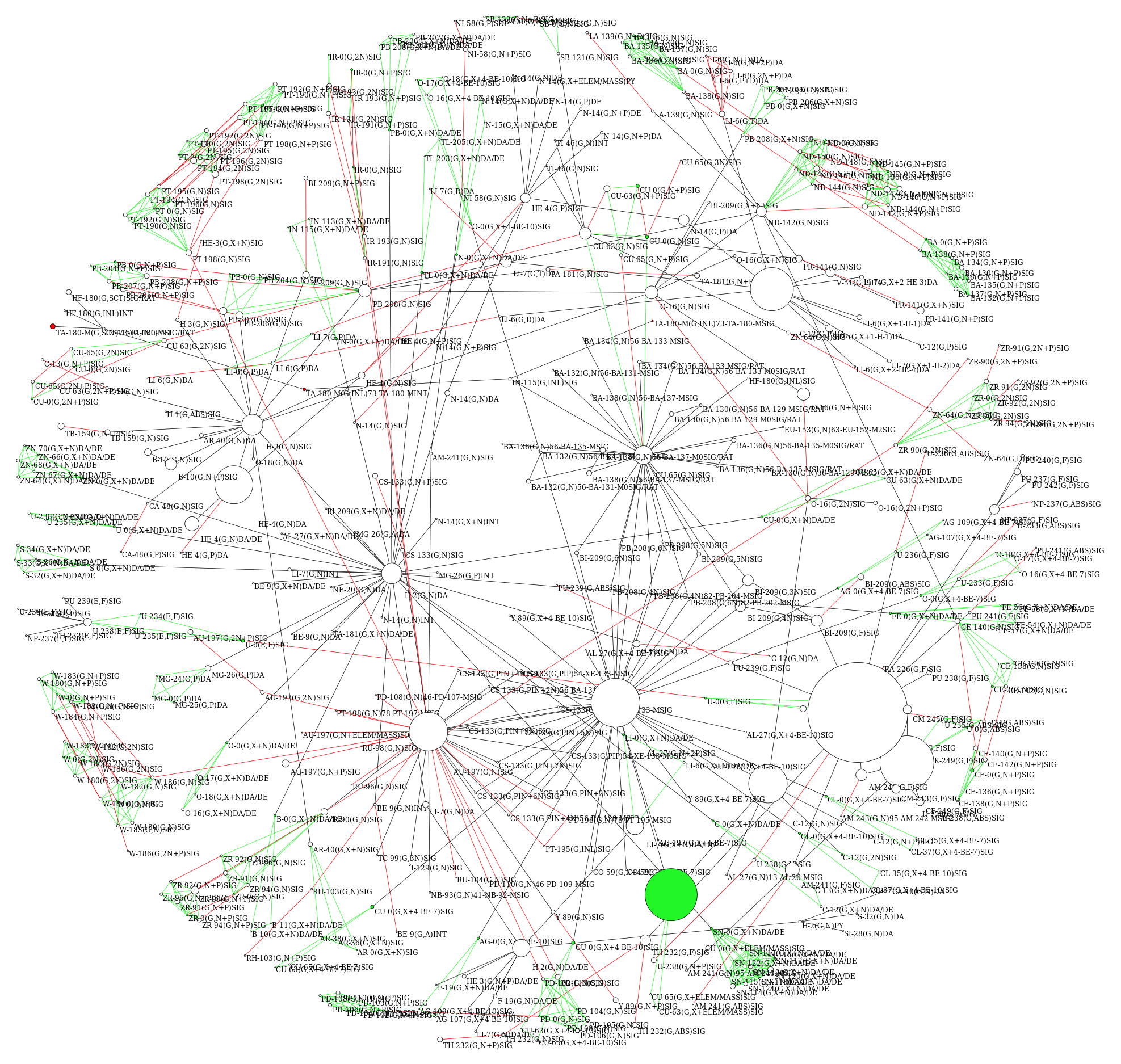}
\caption{\label{fig:sampleGraph}The second largest cluster in the graph.  This cluster is mainly composed of photonuclear data.  The size of the nodes is proportional to the number of occurrences of each reaction/quantity in the EXFOR database.}
\end{figure*}

\begin{figure*}
\includegraphics[width=0.80\textwidth]{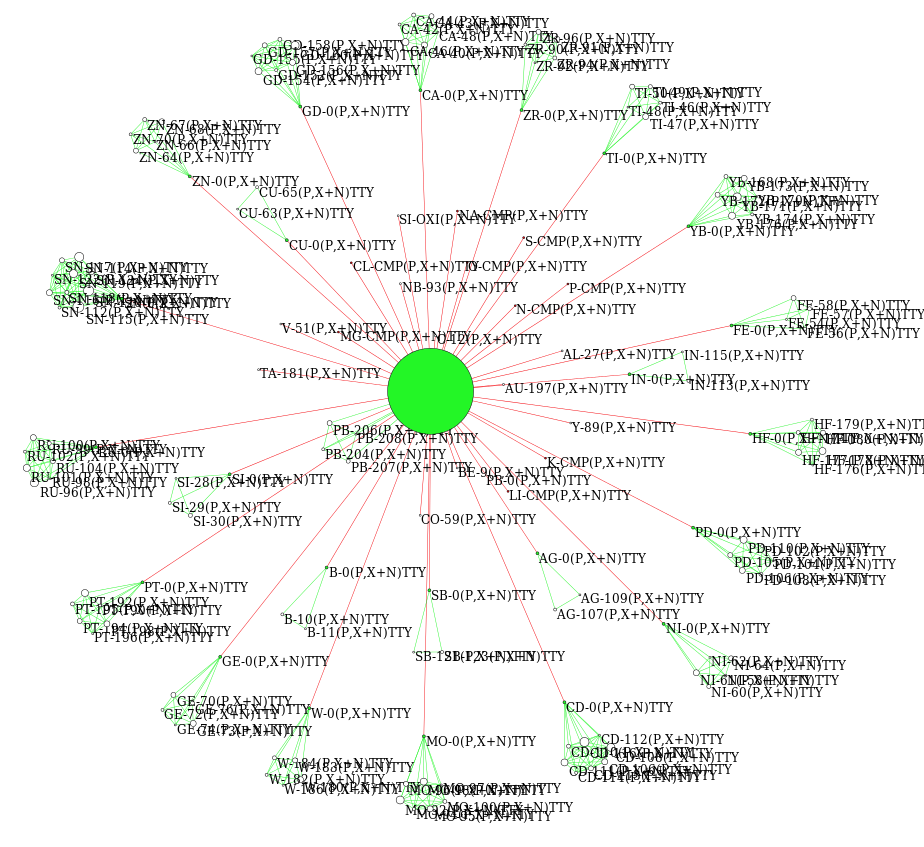}
\caption{\label{fig:sampleGraph2}The third  largest cluster in the graph.  This clusters is mainly composed of TTY  data.  The size of the nodes is proportional to the number of occurrences of each reaction/quantity in the EXFOR database.}
\end{figure*}

\begin{figure*}
\includegraphics[width=0.80\textwidth]{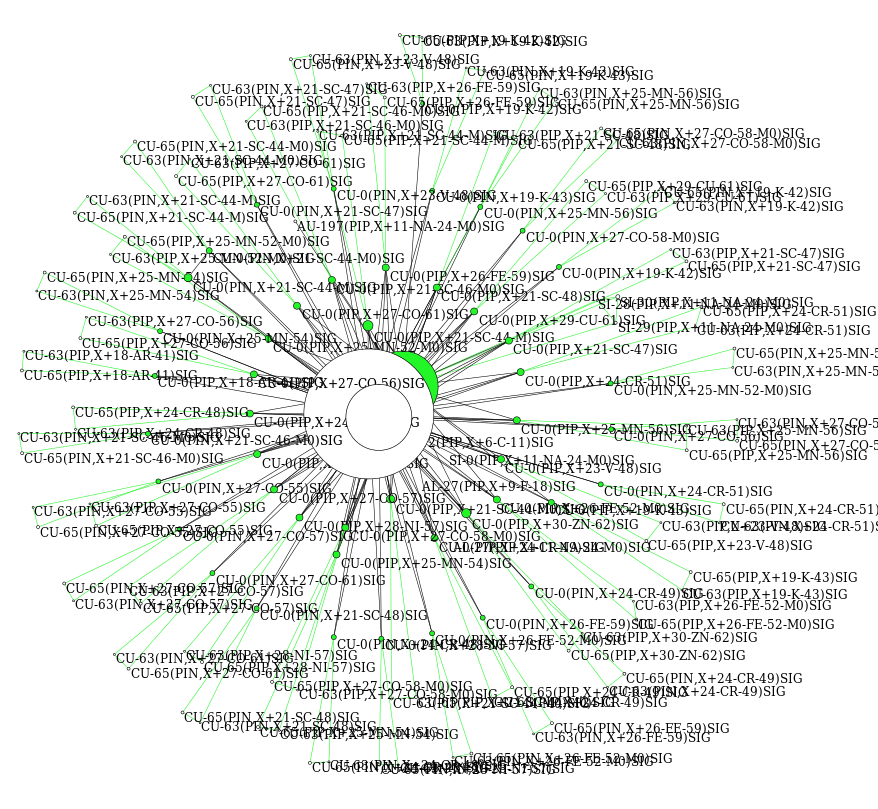}
\caption{\label{fig:sampleGraph3}The  fourth largest cluster in the graph.  This cluster is mainly composed of  charged pion data.  The size of the nodes is proportional to the number of occurrences of each reaction/quantity in the EXFOR database.}
\end{figure*}

We generated our graph using {\tt x4i} \cite{x4i} and saved the results as a GraphML \cite{GraphML} file.  A preliminary versions of this graph were presented in Ref. \cite{brown}.  The full graph has 87,925 nodes and 276,852 edges.   We then studied this graph with the \networkx\, \cite{NetworkX} and \graphtool\, \cite{graphtool} Python packages.  With \graphtool, we were able to visualize portions of the graph. Examples are shown in Figures \ref{fig:sampleGraph}--\ref{fig:sampleGraph3}.   The final graph is too large and fully connected to visualize with the tools we currently have available.  Unfortunately, we were not able to visualize the portion of the graph that contains the majority of the Tier 1 and 2 standards and CIELO nodes.

In the subgraphs in Figures \ref{fig:sampleGraph2} and \ref{fig:sampleGraph3}, one can clearly see the effect of the elemental data: each measurement on a natural target spawns a rosette of isotopic nodes around it.  Adding in the other motifs only increases the local connectedness of the graph.

\begin{figure}
\caption{\label{figure:weightDist}Distribution of the number of occurrences of a reaction/quantity in EXFOR.}
\includegraphics[width=0.45\textwidth]{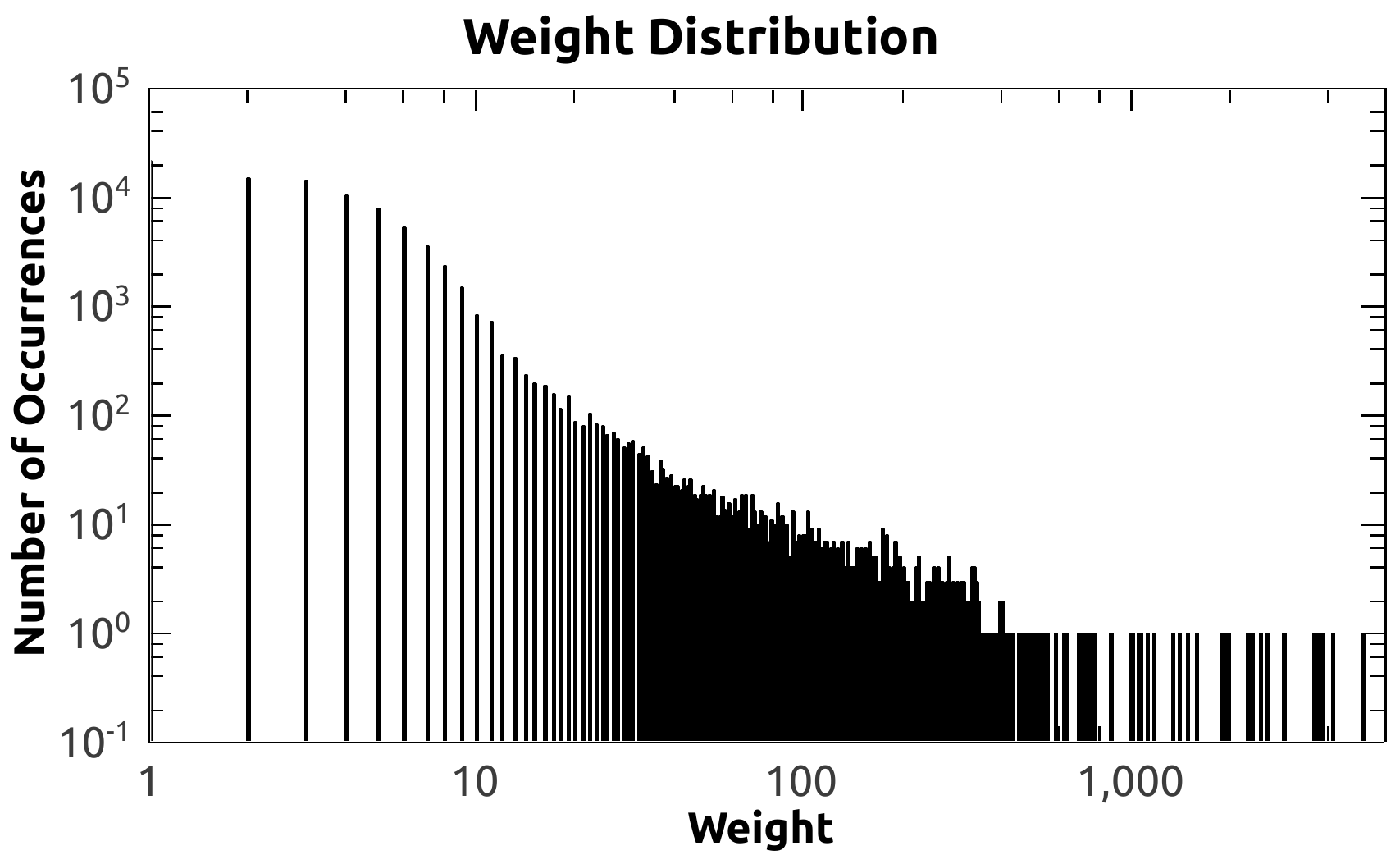}
\end{figure}

In our graph, we rendered the size of a node proportional to the number of occurrences of each reaction/quantity in EXFOR.  Although we understand there are double counting problems in this measure, it is illustrative of the importance of various nodes.  In Figure \ref{figure:weightDist} we present the histogram distribution of the number of occurrences of each reaction/quantity in EXFOR.  There are very few nodes with large weight in this figure and we discuss these in Subsection \ref{section:importByWeight}.  However, there are many nodes with only one measurement.  These are also isolated nodes in that they are not connected to any other node.

We summarize the graph's properties here and discuss them further below:
\begin{itemize}
	\item Number of nodes: 87,925
	\item Number of edges: 276,852
	\item Number of isolates: 23,196
	\item Number of clusters with $40$ or more nodes: 7 
	\item Average degree $\left<k\right>$: 6.2975 
	\item Degree variance $\left<k^2\right>$: 39.6584
	\item Probability two nodes are connected $p$: 7.162e-5
	\item Average cluster coefficient $\left<C\right>$: 0.5958
\end{itemize}

\subsection{Cluster decomposition and the cluster size distribution}

\begin{figure}
\caption{\label{figure:clusterSizeDist}Histogram of disconnected cluster sizes from our graph.}
\includegraphics[width=0.45\textwidth]{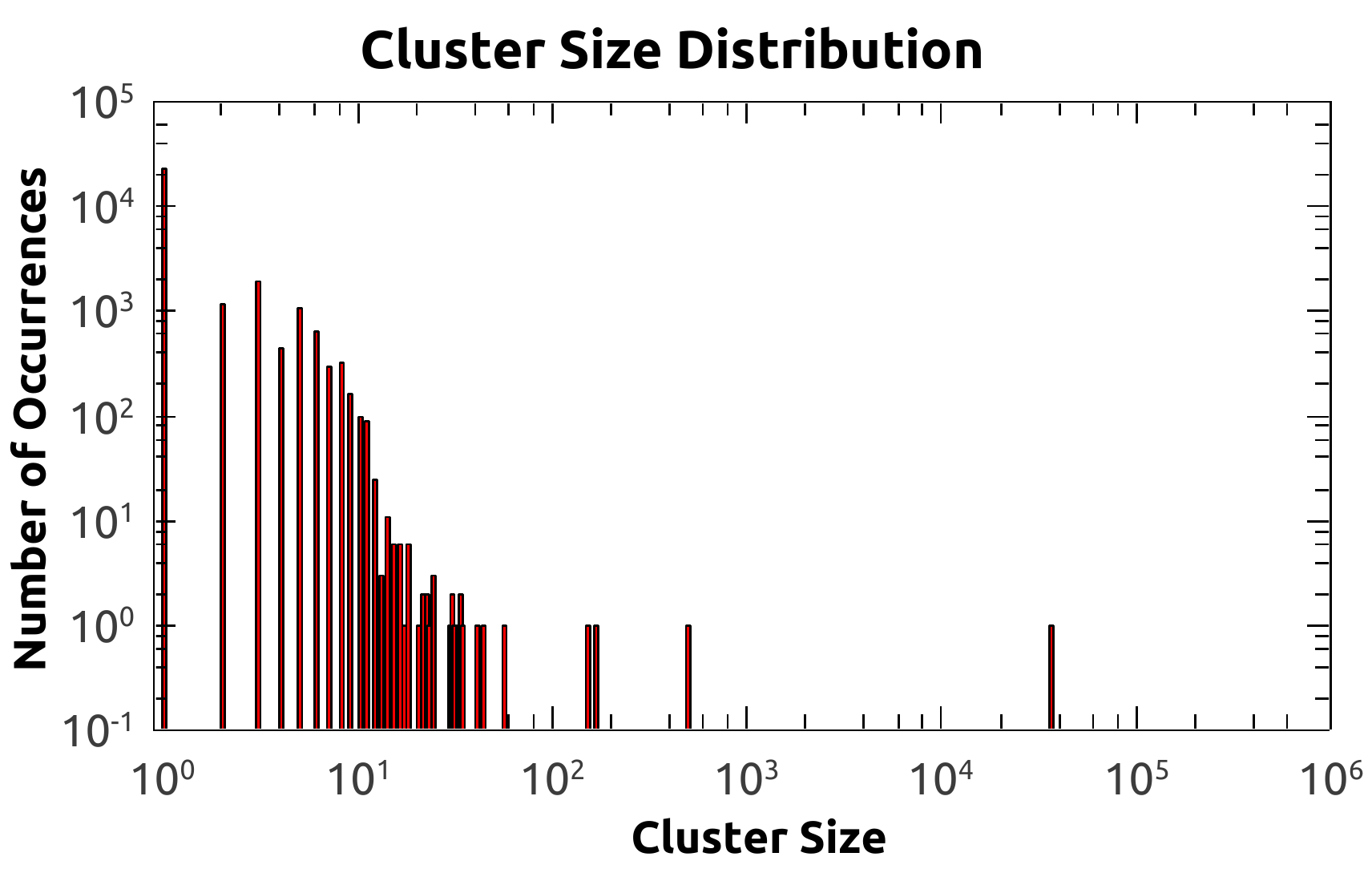}
\end{figure}

When constructed our graph of the EXFOR database, we were surprised to discover that the graph is not fully connected.  Realizing this, we decomposed our graph into disconnected clusters and in Figure \ref{figure:clusterSizeDist} we show the size distribution of the disconnected clusters.  There is one main large cluster containing 35,415 nodes and has less than half the nodes in the graph.  There are thousands of clusters of smaller size and there are 23,196 isolated nodes.  The isolated nodes correspond to experiments that purport to be absolute measurements.  The probability that any two nodes are connected is 7.162e-05. Figure \ref{fig:sampleGraph} is a close-up of the second largest cluster which contains 488 nodes and 963 edges and contains mostly photonuclear data.   Table \ref{table:topClusters} summarizes the content of the largest clusters, containing 40 or more nodes.  In this table, we also list the expected values for the clustering coefficient and average path length for a random graph with the same size as the listed subgraph.

We wonder if all of the disconnected nodes and clusters are as disconnected as our simple analysis implies.  We are limited by the quality of the coding in the EXFOR {\tt ENTRY}s and our ability to parse the EXFOR free text fields.  Also, if the clusters are as disconnected as they appear, is this a good thing?  Quality measurements are often tied to international standards of one form or another.  Nevertheless, the shear number of isolated nodes and small clusters make the prospect of compiling a list of proposed experiments that could someday connect the clusters a daunting task.

\begin{table*}[ht]
\caption{Properties of all clusters containing 40 or more nodes.} 
\centering
\begin{tabular}{lcccccccccc}
\hline\hline
Cluster				 		& $N_{nodes}$	&$N_{edges}$	& $\left<k\right>$	& $\left<k^2\right>$ & $d$  	& $r$ 		& $\ell_{rand}$	& $\ell$ 	& $p=C_{rand}$  		& $\left<C\right>$ \\ \hline\hline
Full graph					& 87,925			& 276,852	& 6.2975				& 39.6584			& n/a	& n/a	 	& n/a			& n/a 	 	& 7.162e-5			& 0.596 \\ 
Main cluster					& 35,415			& 214,300	& 12.1022			& 146.468			& 18		& 10		 	& 4.201			& 5.508  	& 3.417e-4			& 0.758 \\ 
Photonuclear	data				& 488			& 963		& 3.9467				& 15.609				&	18	& 9			& 4.509			& 7.222		& 0.00810			& 0.510\\  
Thick Target Yields (TTY)	    & 164			& 460		& 5.6098				& 31.662				&	4	& 2			& 2.957			& 3.331		& 0.03442			& 0.825\\  
Charged pion reaction data	& 149			& 294		& 3.9463				& 15.679				&	5	& 3			& 3.645			& 3.233		& 0.02666			& 0.782\\  
$p$+Dy data					& 56				& 202		& 7.2143				& 52.992				&	4	& 2 			& 2.037			& 3.151  	& 0.13117			& 0.960 \\   
$^6$Li+Pt data 				& 43				& 132		& 6.1395				& 38.591				&	4	& 2 			& 2.073			& 3.249  	& 0.14618			& 0.939 \\   
$p$+Cs data					& 40				& 87			& 4.35				& 19.408				&	4	& 2 			& 2.509			& 3.109  	& 0.11154			& 0.908 \\   \hline
\end{tabular}
\label{table:topClusters}
\end{table*}

Even though our full graph is disconnected, the individual clusters are connected graphs.  Because each of these clusters is connected, the distance between any two nodes in the cluster is finite.  The distance between any two nodes is defined as the length of the shortest path connecting the two nodes.  If two nodes are not connected, the distance between them is defined as $\infty$.  The average path length of the nodes in a cluster is denoted as $\ell$.  We define the eccentricity of a node as the maximum distance between the node in question and all the other nodes in the graph.  We can compute the diameter of a cluster $d$ as the maximum eccentricity of the nodes in the graph and the radius of a cluster $r$ defined as the minimum eccentricity of the nodes in the graph.  Table \ref{table:topClusters} shows the diameter, radius and path length for the largest clusters.  We note that as the cluster size decreases, the diameter and radius do as well while the clustering coefficient and probability of connection rises.

It is common in network theory to describe a graph as a ``small world'' graph if it meets the following two criteria \cite{AlbertBarabasi}:
\begin{itemize}
\item The average distance is small and typical for a random graph, $\ell \propto \ln(N_{nodes})$.
\item The average clustering coefficient much higher what one expects from a random graph $C_{rand}$ which is just the mean probability of connection $p$.  
\end{itemize}
Here a random graph is one in which one randomly connects nodes with a probability $p$.  By definition, the distance between two disconnected nodes is infinite, therefore our graph as a whole cannot be considered a ``small world'' graph.   However, each of our clusters are ``small world'' graphs in their own right as we see in Table \ref{table:topClusters}.  We exploit this feature later when we provide recommendations for future standards in Section \ref{section:connectToStandards}.

\subsection{Degree distribution}
\begin{figure}
\caption{Degree distribution of reaction/quantity nodes.}
\includegraphics[width=0.45\textwidth]{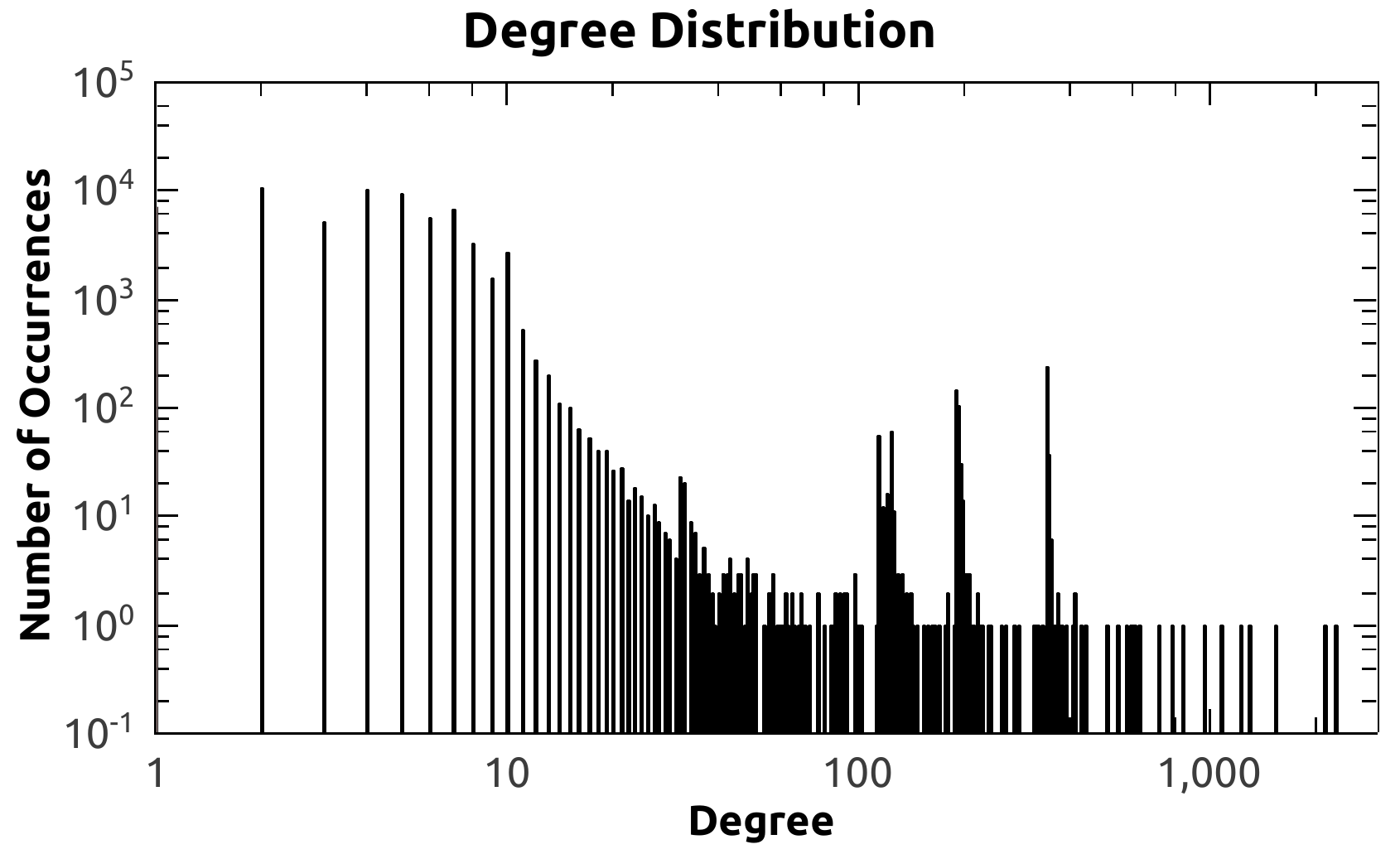}
\label{figure:degreeDist}
\end{figure}

The degree $k$ is simply the number of nodes connecting to a particular node.  It is interesting from a graph theory standpoint to investigate the degree distribution $P(k)$ in that it can help us identify the type of network we have created.  If the degree distribution has a Poisson shape, than it is a random graph.  If on the other hand, it has a power-law fall off, it is likely a scale-free network.  In Figure \ref{figure:degreeDist} we present the degree distribution for our graph.  Our graph clearly is not Poissonian and therefore this strongly suggests that our graph is not a random graph.  While our graph has a plateau from degree 1--8, it seems to exhibit a power-law fall off from roughly degree 10 to degree 80.   The plateau with degree $< 8$ is a typical behavior for many real-world graphs \cite{AlbertBarabasi}.  A power-law ($P(k) \sim k^{-\gamma}$) fit to the degree distribution for $k>5$ yields an exponent of $\gamma=2.5977$, also typical of many real-world networks \cite{AlbertBarabasi}.
 
In Figure \ref{figure:degreeDist}, we note spikes at degrees of 342, 192, 189, 124, 113, 32 and 31.  The spike at a degree of 342 is a cluster of $^{235}$U fission yields.  Similarly, the spikes at 192 and 189 are fission yields for $^{238}$U and $^{239}$Pu respectively.  The spike at 124 is a cluster of $n+^{56}$Fe data and the spike at 113 is a cluster of $n+^{16}$O data.  The spike at 32 seems to be a binning artifact as it it is composed of multiple unrelated nodes with a common degree of 32. Finally, the peak at 31 is comprised of $^1$H reactions.  We note that data for neutron incident reactions on $^1$H, $^{16}$O, $^{56}$Fe,$^{235}$U, $^{238}$U and $^{239}$Pu are all targets of the CIELO evaluation effort \cite{CIELO}.

\subsection{Cluster coefficient distribution}
\begin{figure}
\caption{\label{figure:clusterCoeffDist}Cluster coefficient distribution.}
\includegraphics[width=0.45\textwidth]{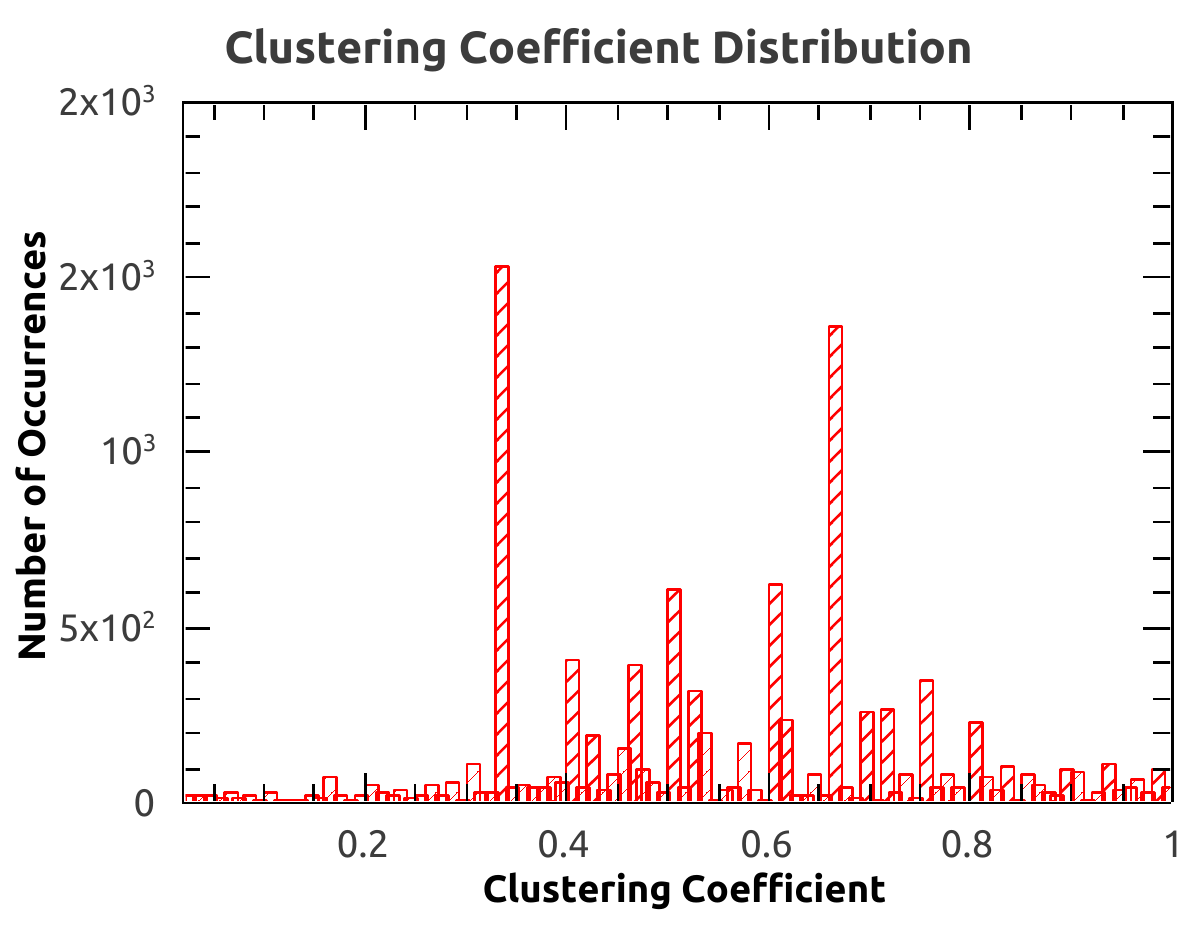}
\end{figure}

The clustering coefficient $C_{i}$ gives the embeddedness of the $i^{th}$ node in the graph in that it tells us how well connected this node is to other nodes in the graph. The $C_i$ is defined as
\begin{equation}
C_{i}=\dfrac{2e_{i}}{k_{i}(k_{i}-1)}
\end{equation}
where $k_i$ is the degree of node $i$ and $e_i$ is the number of edges between the nodes that are directly connected to node $i$.  The maximum value of $e_i$ is $k_i(k_i-1)/2$ so a clustering coefficient of 1 represents a node that is connected to N other nodes,  and those N other nodes are all completely interconnected.  A clustering coefficient of 0 represents an isolated node.  Figure \ref{figure:clusterCoeffDist} shows the distribution of clustering coefficients.  The nodes with the highest clustering coefficient correspond to fission product yields on the major actinides and other related fission data.  These are discussed in Subsection \ref{section:importByClusterCoeff}.  The spikes in the clustering coefficient plot each correspond to elemental clusters and are an artifact of the way in which we added elemental clusters to our graph.

\subsection{Eigenvalue spectrum of the adjacency matrix}
A commonly used metric used to characterized graphs is the eigenvalue spectrum of the graph's adjacency matrix.  The adjacency matrix is a $N_{\rm nodes} \times N_{\rm nodes}$ matrix where $N_{\rm nodes}$ is the number of nodes, in which matrix element $a_{ij}$ is either a one or a zero.  One represents that node $i$ is connected to node $j$, and a zero represents that node $i$ is not connected to node $j$.  For our graph, the adjacency matrix is a $87925 \times 87925$ sparse matrix and is a challenge for most off-the-shelf linear algebra packages.  Were we able to compute the eigenvalue spectrum of our adjacency matrix, we could conclusively state that our graph is not a random graph.  A random graph has a random adjacency matrix.  The hallmark of a random graph is a semicircular eigenvalue spectrum, a.k.a. Wigner's law \cite{AlbertBarabasi}.

\section{Discerning important nodes in the graph}
Our EXFOR network contains nodes ranging from very connected to having no connections at all.  Trying to discern what makes a node/edge important is a matter of opinion and largely depends on the intended use of a reaction/quantity.  In this section we appeal to a variety of graph theory metrics in order to arrive at a list of important nodes.  In the next section, we identify some clearly important nodes and discuss how they can improve the connectivity of our graph.  

\subsection{Ranking by number of occurrences}
\label{section:importByWeight}
There are several ways we could rank the elementary reaction/quantities by importance and the most obvious is just counting the number of occurrences of each reaction/quantity in EXFOR.   One imagines that a node with a large number of occurrences in EXFOR represents large interest from the experimental nuclear science community. Table \ref{table:topByWeight} lists the most important nodes by this weight.

\begin{table*}[ht]
\caption{\label{table:topByWeight}Top 50 nodes ranked by number of occurrences of a reaction/quantity in EXFOR.   References marked with a ``*'' indicate that the reaction/quantity are a by-product of the Neutron Standards Project evaluation process.} 
\begin{tabular}{llcl}
\hline\hline
Name & Observable & \# Occurances & Reference\\
\hline\hline
$^{27}$Al($n$, $\alpha$) & $\sigma$ & 5049 & Proposed \cite{IAEAStdsMeeting2013}, IRDFF \cite{IRDFF}\\
$^{197}$Au($n$, $\gamma$) & $\sigma$ & 4106 & ENDF/B-VII.1 Standard \cite{Standards}, Atlas \cite{Atlas}, IRDFF \cite{IRDFF}\\
$^{27}$Al($p$, X+$^{22}$Na) & $\sigma$ & 3806 & Medical/Dosimeter \cite{IAEAMedical}\\
$^{235}$U($n$, f) & $\sigma$ & 3707 & ENDF/B-VII.1 Standard \cite{Standards}, CIELO \cite{CIELO}\\
$^{27}$Al($p$, X+$^{24}$Na) & $\sigma$ & 3626 & Medical/Dosimeter \cite{IAEAMedical}\\
$^{1}$H($n$, el) & $\sigma$ & 2903 & ENDF/B-VII.1 Standard \cite{Standards}, CIELO \cite{CIELO}\\
$^{1}$H($n$, el) & $d\sigma/d\Omega$ & 2601 & CIELO \cite{CIELO}\\
$^{93}$Nb($n$, $2n$+$^{92m}$Nb) & $\sigma$ & 2465 & IRDFF \cite{IRDFF}\\
$^{27}$Al($p$, $n$+$3p$) & $\sigma$ & 2316 & \\
$^{56}$Fe($n$, $p$+$^{56}$Mn) & $\sigma$ & 2272 & CIELO \cite{CIELO}\\
$^{197}$Au($n$, $\gamma$) & RI & 1961 & Atlas \cite{Atlas}\\
$^{27}$Al($n$, $p$+$^{27}$Mg) & $\sigma$ & 1902 & \\
$^{nat}$Cu($p$, X+$^{65}$Zn) & $\sigma$ & 1899 & Medical/Dosimeter \cite{IAEAMedical}\\
$^{59}$Co($n$, $\gamma$) & RI & 1582 & Atlas \cite{Atlas}\\
$^{58}$Ni($n$, $p$) & $\sigma$ & 1477 & IRDFF \cite{IRDFF}\\
$^{238}$U($n$, f) & $\sigma$ & 1394 & ENDF/B-VII.1 Standard \cite{Standards}, CIELO \cite{CIELO}\\
$^{59}$Co($n$, $\gamma$) & $\sigma$ & 1332 & Atlas \cite{Atlas}, IRDFF \cite{IRDFF}\\
$^{115}$In($n$, inel) & $\sigma$ & 1161 & IRDFF \cite{IRDFF}\\
$^{nat}$Mo($p$, X+$^{96}$Tc) & $\sigma$ & 1109 & \\
$^{27}$Al($^{12}$C, X+$^{24}$Na) & $\sigma$ & 1060 & \\
$^{235}$U($n$, f) & FPY (ELEM/MASS) & 1059 & IRDFF \cite{IRDFF}\\
$^{238}$U($n$, $\gamma$) & RI & 1056 & CIELO \cite{CIELO}\\
$^{239}$Pu($n$, f) & $\sigma$ & 1004 & ENDF/B-VII.1 Standard$^*$ \cite{Standards}, CIELO \cite{CIELO}\\
$^{238}$U($n$, $\gamma$) & $\sigma$ & 1003 & ENDF/B-VII.1 Standard$^*$ \cite{Standards}, IRDFF \cite{IRDFF}, CIELO \cite{CIELO}\\
$^{27}$Al($d$, X+$^{24}$Na) & $\sigma$ & 990 & Medical/Dosimeter \cite{IAEAMedical}\\
$^{nat}$Cu($p$, X+$^{62}$Zn) & $\sigma$ & 985 & Medical/Dosimeter \cite{IAEAMedical}\\
$^{10}$B($n$, $\alpha$) & $\sigma$ & 860 & ENDF/B-VII.1 Standard \cite{Standards}, IRDFF \cite{IRDFF}\\
$^{6}$Li($n$, $t$) & $\sigma$ & 761 & ENDF/B-VII.1 Standard \cite{Standards}, IRDFF \cite{IRDFF}\\
$^{63}$Cu($n$, $2n$) & $\sigma$ & 752 & IRDFF \cite{IRDFF}\\
$^{235}$U($n$, f) & $\bar{\nu}$ & 740 & CIELO \cite{CIELO}\\
$^{nat}$Ti($p$, X+$^{48}$V) & $\sigma$ & 731 & Medical/Dosimeter \cite{IAEAMedical}\\
$^{nat}$Ti($d$, X+$^{48}$V) & $\sigma$ & 713 & Medical/Dosimeter \cite{IAEAMedical}\\
$^{56}$Fe($n$, inel) & $\sigma$ & 694 & CIELO \cite{CIELO}\\
$^{127}$I($n$, $\gamma$) & $\sigma$ & 633 & \\
$^{45}$Sc($n$, $\gamma$) & RI & 617 & \\
$^{235}$U($n$, f) & FPY & 586 & \\
$^{nat}$Mo($\alpha$, X+$^{97}$Ru) & $\sigma$ & 547 & \\
$^{54}$Fe($n$, $p$+$^{54}$Mn) & $\sigma$ & 541 & \\
$^{27}$Al($p$, $3n$+$3p$) & $\sigma$ & 528 & \\
$^{235}$U($n$, f) & FPY (MASS) & 521 & \\
$^{65}$Cu($p$, $n$) & $\sigma$ & 514 & \\
$^{60}$Ni($n$, $p$) & $\sigma$ & 496 & IRDFF \cite{IRDFF}\\
$^{235}$U($n$, f+$^{99}$Mo) & FPY & 489 & \\
$^{65}$Cu($n$, $2n$) & $\sigma$ & 482 & IRDFF \cite{IRDFF}\\
$^{239}$Pu($n$, f) & $\bar{\nu}$ & 470 & CIELO \cite{CIELO}\\
$^{12}$C($n$, el) & $\sigma$ & 455 & \\
$^{55}$Mn($n$, $\gamma$+$^{56}$Mn) & $\sigma$ & 452 & \\
$^{27}$Al($p$, X+$^{7}$Be) & $\sigma$ & 434 & \\
$^{252}$Cf($0$, f) & $\bar{\nu}$ & 432 & Proposed \cite{IAEAStdsMeeting2013}\\
$^{56}$Fe($n$, inel) & $d\sigma/d\Omega$ & 422 & CIELO \cite{CIELO}\\
\hline
\end{tabular}
\end{table*}

We note that while the Tier 1 standards are well represented in this table, they are not uniformly the most highly ranked by this measure.  The Tier 2 standards are also represented well, but there are several nodes that are not viewed as important by any standards project.  Various reactions on aluminum are clearly important and this is no surprise as aluminum is readily available, durable and mono-isotopic making it an ideal target material.  We see also reactions on molybdenum are important, most likely for medical isotope production.  Finally, some fission product yield (FPY) data has a relatively high ranking.

\subsection{Ranking by degree}
\label{section:importByDegree}
We could also rank the reaction/quantities by number of connections to other reactions/quantities.  Nodes with a large degree may be commonly used as reaction monitors.  In Table \ref{table:topByDegree}, we list the most important nodes rank ordered by degree.  Interestingly, the top four nodes by degree or number of occurrences are the same, although the rank order is different.  Also interesting is the fact that five of the top twenty reactions/quantities are not addressed by any standards effort

\begin{table*}[ht]
\caption{\label{table:topByDegree}Top 50 nodes ranked by node degree.  References marked with a ``*'' indicate that the reaction/quantity are a by-product of the Neutron Standards Project evaluation process.} 
\begin{tabular}{llcl}
\hline\hline
Name & Observable & Degree & Reference\\
\hline\hline
$^{27}$Al($p$, X+$^{22}$Na) & $\sigma$ & 2274 & Medical/Dosimeter \cite{IAEAMedical}\\
$^{27}$Al($p$, X+$^{24}$Na) & $\sigma$ & 2120 & Medical/Dosimeter \cite{IAEAMedical}\\
$^{27}$Al($p$, $n$+$3p$) & $\sigma$ & 1533 & \\
$^{27}$Al($n$, $\alpha$) & $\sigma$ & 1279 & Proposed \cite{IAEAStdsMeeting2013}, IRDFF \cite{IRDFF}\\
$^{1}$H($n$, el) & $\sigma$ & 1207 & ENDF/B-VII.1 Standard \cite{Standards}, CIELO \cite{CIELO}\\
$^{197}$Au($n$, $\gamma$) & $\sigma$ & 1071 & ENDF/B-VII.1 Standard \cite{Standards}, Atlas \cite{Atlas}, IRDFF \cite{IRDFF}\\
$^{1}$H($n$, el) & $d\sigma/d\Omega$ & 951 & CIELO \cite{CIELO}\\
$^{56}$Fe($n$, $p$+$^{56}$Mn) & $\sigma$ & 831 & CIELO \cite{CIELO}\\
$^{235}$U($n$, f) & $\sigma$ & 772 & ENDF/B-VII.1 Standard \cite{Standards}, CIELO \cite{CIELO}\\
$^{93}$Nb($n$, $2n$+$^{92m}$Nb) & $\sigma$ & 708 & IRDFF \cite{IRDFF}\\
$^{nat}$Cu($p$, X+$^{65}$Zn) & $\sigma$ & 625 & Medical/Dosimeter \cite{IAEAMedical}\\
$^{27}$Al($^{12}$C, X+$^{24}$Na) & $\sigma$ & 610 & \\
$^{nat}$Mo($p$, X+$^{96}$Tc) & $\sigma$ & 600 & \\
$^{nat}$Mo($\alpha$, X+$^{97}$Ru) & $\sigma$ & 594 & \\
$^{59}$Co($n$, $\gamma$) & $\sigma$ & 576 & Atlas \cite{Atlas}, IRDFF \cite{IRDFF}\\
$^{27}$Al($n$, $p$+$^{27}$Mg) & $\sigma$ & 542 & \\
$^{238}$U($n$, f) & $\sigma$ & 509 & ENDF/B-VII.1 Standard \cite{Standards}, CIELO \cite{CIELO}\\
$^{27}$Al($d$, X+$^{24}$Na) & $\sigma$ & 507 & Medical/Dosimeter \cite{IAEAMedical}\\
$^{197}$Au($n$, $\gamma$) & RI & 438 & Atlas \cite{Atlas}\\
$^{10}$B($n$, $\alpha$) & $\sigma$ & 430 & ENDF/B-VII.1 Standard \cite{Standards}, IRDFF \cite{IRDFF}\\
$^{235}$U($n$, f) & RI & 425 & CIELO \cite{CIELO}\\
$^{nat}$Cu($p$, X+$^{62}$Zn) & $\sigma$ & 411 & Medical/Dosimeter \cite{IAEAMedical}\\
$^{59}$Co($n$, $\gamma$) & RI & 408 & Atlas \cite{Atlas}\\
$^{235}$U($n$, f) & $\bar{\nu}$ & 408 & CIELO \cite{CIELO}\\
$^{235}$U($n$, f) & FPY (ELEM/MASS) & 406 & IRDFF \cite{IRDFF}\\
$^{235}$U($n$, abs) & $\sigma$ & 386 & CIELO \cite{CIELO}\\
$^{6}$Li($n$, $t$) & $\sigma$ & 379 & ENDF/B-VII.1 Standard \cite{Standards}, IRDFF \cite{IRDFF}\\
$^{235}$U($n$, f+$^{99}$Mo) & FPY & 375 & \\
$^{238}$U($n$, $\gamma$) & $\sigma$ & 373 & ENDF/B-VII.1 Standard$^*$ \cite{Standards}, IRDFF \cite{IRDFF}, CIELO \cite{CIELO}\\
$^{235}$U($n$, f) & AKE & 366 & CIELO \cite{CIELO}\\
$^{235}$U($n$, $\gamma$) & $\sigma$ & 366 & CIELO \cite{CIELO}\\
$^{235}$U($n$, f+$^{140}$Ba) & FPY & 360 & \\
$^{235}$U($n$, f) & FPY & 358 & \\
$^{235}$U($n$, el) & $\sigma$ & 353 & CIELO \cite{CIELO}\\
$^{235}$U($n$, el) & $d\sigma/d\Omega$ & 352 & CIELO \cite{CIELO}\\
$^{235}$U($n$, f) & FPY (MASS) & 352 & \\
$^{235}$U($n$, f) & KE & 352 & \\
$^{235}$U($n$, abs) & ETA & 351 & CIELO \cite{CIELO}\\
$^{235}$U($n$, f+$^{140}$La) & $\sigma$ & 350 & CIELO \cite{CIELO}\\
$^{235}$U($n$, tot) & $\sigma$ & 350 & CIELO \cite{CIELO}\\
$^{235}$U($n$, f+$n$) & KE & 350 & \\
$^{235}$U($n$, inel) & $\sigma$ & 349 & CIELO \cite{CIELO}\\
$^{235}$U($n$, f+$^{145}$Nd) & FPY & 349 & \\
$^{235}$U($n$, f) & $d\nu/dE'$ & 349 & CIELO \cite{CIELO}\\
$^{235}$U($n$, abs) & ALF & 349 & CIELO \cite{CIELO}\\
$^{235}$U($n$, f) & INT & 349 & CIELO \cite{CIELO}\\
$^{63}$Cu($n$, $2n$) & $\sigma$ & 349 & IRDFF \cite{IRDFF}\\
$^{235}$U($n$, f) & ARE & 348 & CIELO \cite{CIELO}\\
$^{235}$U($n$, f+$^{136}$Cs) & FPY & 348 & \\
$^{235}$U($n$, inel) & $d\sigma/d\Omega$ & 348 & CIELO \cite{CIELO}\\
\hline
\end{tabular}
\end{table*}

As one can see from Table \ref{table:topByDegree}, quite a few nodes that have a high degree and are not considered either Tier 1 or 2 standards. As in the previous case, reactions on aluminum and molybdenum rank highly.  Towards the bottom of this table, we begin to see fission product yields.

\subsection{Ranking by cluster coefficient}
\label{section:importByClusterCoeff}

In Table \ref{table:topByClusterCoeff}, we list the top nodes ranked by largest cluster coefficient.  These are the nodes that are in the most highly interconnected regions of the main cluster.  The list is entirely comprised of data on the three major actinides ($^{235}$U, $^{238}$U, and $^{239}$Pu).  Most of the entries on this list are cross section and related data covered by the CIELO pilot project \cite{CIELO}.  The rest are various fission product yield data.  Generally when one measures fission product yields, one does it for many reaction products at once, introducing a large number of connections in our graph.  This also accounts for the fact several nodes have identical clustering coefficients, implying that they are all part of the same grouping of nodes.

\begin{table*}[ht]
\caption{\label{table:topByClusterCoeff}Top 50 nodes ranked by cluster coefficient.} 
\begin{tabular}{llcl}
\hline\hline
Name & Observable & Cluster coefficient & Reference\\
\hline\hline
$^{235}$U($n$, f+$^{139}$Ba) & FPY & 0.994390738751641 & \\
$^{235}$U($n$, f+$^{137}$Cs) & FPY & 0.994390738751641 & \\
$^{235}$U($n$, f+$^{135}$Cs) & FPY & 0.994390738751641 & \\
$^{235}$U($n$, f+$^{134m}$Sb) & FPY & 0.994390738751641 & \\
$^{235}$U($n$, f+$^{121m}$In) & FPY & 0.994390738751641 & \\
$^{235}$U($n$, non) & DE & 0.994356639899067 & CIELO \cite{CIELO}\\
$^{235}$U($n$, inel) & KE & 0.994356639899067 & CIELO \cite{CIELO}\\
$^{235}$U($n$, inel) & DE & 0.994356639899067 & CIELO \cite{CIELO}\\
$^{235}$U($n$, X+$n$) & $\sigma$ & 0.994237293915059 & CIELO \cite{CIELO}\\
$^{235}$U($n$, f+$^{131}$Xe) & FPY & 0.994237293915059 & \\
$^{235}$U($n$, f+$^{105}$Pd) & FPY & 0.994237293915059 & \\
$^{235}$U($n$, f) & DA/DE & 0.994237293915059 & CIELO \cite{CIELO}\\
$^{235}$U($n$, $3n$) & $\sigma$ & 0.994237293915059 & CIELO \cite{CIELO}\\
$^{235}$U($n$, $0$) & J & 0.994237293915059 & CIELO \cite{CIELO}\\
$^{235}$U($n$, X+$\gamma$) & $\sigma$ & 0.994220244488773 & CIELO \cite{CIELO}\\
$^{235}$U($n$, X+$\gamma$) & DE & 0.994220244488773 & CIELO \cite{CIELO}\\
$^{235}$U($n$, f) & SPC (ELEM/MASS) & 0.994220244488773 & \\
$^{235}$U($n$, f) & DE (MASS) & 0.994203195062486 & \\
$^{235}$U($n$, f) & DE & 0.994203195062486 & CIELO \cite{CIELO}\\
$^{235}$U($n$, f+$^{91}$Sr) & FPY & 0.994186145636199 & \\
$^{235}$U($n$, $p$) & $\sigma$ & 0.994169096209912 & CIELO \cite{CIELO}\\
$^{235}$U($n$, f) & FPY (ELEM) & 0.994169096209912 & \\
$^{235}$U($n$, f+$^{133}$Sb) & FPY & 0.994169096209912 & \\
$^{235}$U($n$, f+$^{132}$Sb) & FPY & 0.994169096209912 & \\
$^{235}$U($n$, f) & WID & 0.994169096209912 & CIELO \cite{CIELO}\\
$^{235}$U($n$, f) & SIG/RAT & 0.994169096209912 & CIELO \cite{CIELO}\\
$^{238}$U($n$, inel) & DE & 0.990446891191709 & CIELO \cite{CIELO}\\
$^{238}$U($n$, inel) & DA/DE & 0.990446891191709 & CIELO \cite{CIELO}\\
$^{238}$U($n$, f) & AKE/DA & 0.990446891191709 & CIELO \cite{CIELO}\\
$^{238}$U($n$, f) & $d\sigma/d\Omega$ (MASS) & 0.990284974093264 & \\
$^{238}$U($n$, f) & RI & 0.990284974093264 & CIELO \cite{CIELO}\\
$^{238}$U($n$, abs) & ETA & 0.990284974093264 & CIELO \cite{CIELO}\\
$^{239}$Pu($n$, tot) & $\sigma$ & 0.990086326928432 & CIELO \cite{CIELO}\\
$^{239}$Pu($n$, sct) & $\sigma$ & 0.990086326928432 & CIELO \cite{CIELO}\\
$^{238}$U($n$, el) & POT & 0.990069084628670 & CIELO \cite{CIELO}\\
$^{238}$U($n$, $\gamma$) & SGV & 0.990015112262521 & CIELO \cite{CIELO}\\
$^{238}$U($n$, $\gamma$) & MLT & 0.990015112262521 & CIELO \cite{CIELO}\\
$^{238}$U($n$, $\gamma$) & DA/DE & 0.990015112262521 & CIELO \cite{CIELO}\\
$^{238}$U($n$, f) & SGV & 0.990015112262521 & CIELO \cite{CIELO}\\
$^{238}$U($n$, $2n$) & SGV & 0.990015112262521 & CIELO \cite{CIELO}\\
$^{238}$U($n$, f+$n$) & PR & 0.989961139896373 & CIELO \cite{CIELO}\\
$^{238}$U($n$, f) & PR & 0.989961139896373 & CIELO \cite{CIELO}\\
$^{239}$Pu($n$, non) & DE & 0.989863547758284 & CIELO \cite{CIELO}\\
$^{239}$Pu($n$, inel) & DE & 0.989863547758284 & CIELO \cite{CIELO}\\
$^{238}$U($n$, X+$\gamma$) & DE & 0.989853195164076 & CIELO \cite{CIELO}\\
$^{238}$U($n$, f) & SPC (ELEM/MASS) & 0.989853195164076 & \\
$^{238}$U($n$, f+$^{4}$He) & FPY & 0.989799222797927 & \\
$^{239}$Pu($n$, f+$^{97}$Zr) & FPY & 0.989640768588137 & \\
$^{239}$Pu($n$, $3n$) & $\sigma$ & 0.989640768588137 & CIELO \cite{CIELO}\\
\hline
\end{tabular}
\end{table*}

\subsection{Centrality/importance measures that depend on the adjacency matrix}
We attempted to use several other measures of node importance including node centrality, betweenness and eigenvalue centrality \cite{NetworkX,graphtool}.  All of these measures fail for our graph because the graph is too large and these measures rely on performing complex linear algebra on the adjacency matrix of the graph.

\subsection{Ranking by PageRank}
\label{section:importByPageRank}
There is one widely used measure which does not rely on the adjacency matrix: Google's PageRank \cite{PageRank}.  PageRank is an iterative process to determine what the probability is that a given node is connected.  The exact algorithm is given in many places and is implemented in the codes we used for our analysis (see reference \cite{PageRank}).   This algorithm is robust and simple and can work on graphs as large as the entire Internet in a reasonable amount of time.  Table \ref{table:topByPageRank} lists the top nodes ranked by PageRank.  Using PageRank, we again find reactions on aluminum and molybdenum rank highly.

\begin{table*}[ht]
\caption{\label{table:topByPageRank}Top 50 nodes ranked by Google PageRank.} 
\begin{tabular}{llcl}
\hline\hline
Name & Observable & PageRank & Reference\\
\hline\hline
$^{27}$Al($p$, X+$^{22}$Na) & $\sigma$ & 0.00616981360379 & Medical/Dosimeter \cite{IAEAMedical}\\
$^{27}$Al($p$, X+$^{24}$Na) & $\sigma$ & 0.00592690853695 & Medical/Dosimeter \cite{IAEAMedical}\\
$^{27}$Al($p$, $n$+$3p$) & $\sigma$ & 0.00464164933365 & \\
$^{1}$H($n$, el) & $\sigma$ & 0.00206875556656 & ENDF/B-VII.1 Standard \cite{Standards}, CIELO \cite{CIELO}\\
$^{27}$Al($n$, $\alpha$) & $\sigma$ & 0.00202782968132 & Proposed \cite{IAEAStdsMeeting2013}, IRDFF \cite{IRDFF}\\
$^{27}$Al($^{12}$C, X+$^{24}$Na) & $\sigma$ & 0.00199257000533 & \\
$^{1}$H($n$, el) & $d\sigma/d\Omega$ & 0.0016782550455 & CIELO \cite{CIELO}\\
$^{197}$Au($n$, $\gamma$) & $\sigma$ & 0.00163673076474 & ENDF/B-VII.1 Standard \cite{Standards}, Atlas \cite{Atlas}, IRDFF \cite{IRDFF}\\
$^{nat}$Mo($p$, X+$^{96}$Tc) & $\sigma$ & 0.00152614933497 & \\
$^{nat}$Mo($\alpha$, X+$^{97}$Ru) & $\sigma$ & 0.00151440986728 & \\
$^{nat}$Cu($p$, X+$^{65}$Zn) & $\sigma$ & 0.00139712797412 & Medical/Dosimeter \cite{IAEAMedical}\\
$^{27}$Al($d$, X+$^{24}$Na) & $\sigma$ & 0.00126335663443 & Medical/Dosimeter \cite{IAEAMedical}\\
$^{56}$Fe($n$, $p$+$^{56}$Mn) & $\sigma$ & 0.00112176853498 & CIELO \cite{CIELO}\\
$^{93}$Nb($n$, $2n$+$^{92m}$Nb) & $\sigma$ & 0.00105648813828 & IRDFF \cite{IRDFF}\\
$^{65}$Cu($p$, $n$) & $\sigma$ & 0.000855440400277 & \\
$^{59}$Co($n$, $\gamma$) & $\sigma$ & 0.00083651416181 & Atlas \cite{Atlas}, IRDFF \cite{IRDFF}\\
$^{nat}$Cu($p$, X+$^{62}$Zn) & $\sigma$ & 0.000818153268765 & Medical/Dosimeter \cite{IAEAMedical}\\
$^{27}$Al($n$, $p$+$^{27}$Mg) & $\sigma$ & 0.000766232991113 & \\
$^{27}$Al($p$, $3n$+$3p$) & $\sigma$ & 0.00073430981797 & \\
$^{nat}$Ti($p$, X+$^{48}$V) & $\sigma$ & 0.000715429041296 & Medical/Dosimeter \cite{IAEAMedical}\\
$^{nat}$Ti($d$, X+$^{48}$V) & $\sigma$ & 0.000700853239592 & Medical/Dosimeter \cite{IAEAMedical}\\
$^{10}$B($n$, $\alpha$) & $\sigma$ & 0.00069397317621 & ENDF/B-VII.1 Standard \cite{Standards}, IRDFF \cite{IRDFF}\\
$^{27}$Al($p$, X+$^{7}$Be) & $\sigma$ & 0.000633605482474 & \\
$^{6}$Li($n$, $t$) & $\sigma$ & 0.000589643577153 & ENDF/B-VII.1 Standard \cite{Standards}, IRDFF \cite{IRDFF}\\
$^{235}$U($n$, f) & $\sigma$ & 0.000589527515957 & ENDF/B-VII.1 Standard \cite{Standards}, CIELO \cite{CIELO}\\
$^{65}$Cu($p$, X+$^{64}$Cu) & $\sigma$ & 0.000581555115186 & \\
$^{27}$Al($d$, X+$^{22}$Na) & $\sigma$ & 0.000532941906423 & Medical/Dosimeter \cite{IAEAMedical}\\
$^{197}$Au($n$, $\gamma$) & RI & 0.000527576761568 & Atlas \cite{Atlas}\\
$^{27}$Al($\alpha$, X+$^{24}$Na) & $\sigma$ & 0.000526693344035 & Medical/Dosimeter \cite{IAEAMedical}\\
$^{59}$Co($n$, $\gamma$) & RI & 0.00052515716658 & Atlas \cite{Atlas}\\
$^{65}$Cu($\alpha$, $2n$) & $\sigma$ & 0.000501957045702 & \\
$^{63}$Cu($n$, $2n$) & $\sigma$ & 0.000499570466347 & IRDFF \cite{IRDFF}\\
$^{63}$Cu($p$, $n$) & $\sigma$ & 0.000487493703315 & \\
$^{238}$U($n$, f) & $\sigma$ & 0.000459618986486 & ENDF/B-VII.1 Standard \cite{Standards}, CIELO \cite{CIELO}\\
$^{12}$C($d$, X+$^{11}$C) & $\sigma$ & 0.000458547158062 & \\
$^{nat}$Ti($^{3}$He, X+$^{48}$V) & $\sigma$ & 0.000420732688038 & Medical/Dosimeter \cite{IAEAMedical}\\
$^{58}$Ni($n$, $p$) & $\sigma$ & 0.000419142073153 & IRDFF \cite{IRDFF}\\
$^{27}$Al($d$, $p$+$\alpha$) & $\sigma$ & 0.000416101528133 & \\
$^{27}$Al($\alpha$, X+$^{22}$Na) & $\sigma$ & 0.000336414603953 & Medical/Dosimeter \cite{IAEAMedical}\\
$^{nat}$C($n$, el) & $\sigma$ & 0.000332529924528 & ENDF/B-VII.1 Standard \cite{Standards}\\
$^{75}$As($n$, $2n$) & $\sigma$ & 0.000318043040825 & IRDFF \cite{IRDFF}\\
$^{127}$I($n$, $\gamma$) & $\sigma$ & 0.000313626212101 & \\
$^{3}$H($d$, $n$) & $\sigma$ & 0.000307358539955 & \\
$^{nat}$Cu($p$, X+$^{63}$Zn) & $\sigma$ & 0.000301789432478 & Medical/Dosimeter \cite{IAEAMedical}\\
$^{65}$Cu($n$, $2n$) & $\sigma$ & 0.000294059042661 & IRDFF \cite{IRDFF}\\
$^{63}$Cu($p$, $2n$) & $\sigma$ & 0.000293541428666 & \\
$^{nat}$Ti($\alpha$, X+$^{51}$Cr) & $\sigma$ & 0.000292359776106 & \\
$^{54}$Fe($n$, $p$+$^{54}$Mn) & $\sigma$ & 0.000283550536805 & \\
$^{nat}$C($n$, el) & $d\sigma/d\Omega$ & 0.000279013304815 & \\
$^{115}$In($n$, inel) & $\sigma$ & 0.000274291367779 & IRDFF \cite{IRDFF}\\
\hline
\end{tabular}
\end{table*}

\section{Characterizing the connectivity to standards}
\label{section:connectToStandards}
It is clear from the analysis above that the following reaction/quantities have out-sized importance as measured by several different metrics: 
\begin{itemize}
\item Aluminum reaction/quantities:
\begin{itemize}
\item n+$^{27}$Al: the ($n,p+^{27}$Mg) cross section
\item p+$^{27}$Al: the ($p,n+3p$) cross section and the $^{22}$Na and $^{24}$Na production cross sections
\item $^{12}$C+$^{27}$Al: the $^{24}$Na production cross section
\end{itemize}
\item Molybdinum is also a very important structural material:
\begin{itemize}
\item p+$^{nat}$Mo: the $^{96}$Tc production cross section
\item $\alpha$ + $^{nat}$Mo: the $^{97}$Ru production cross section
\end{itemize}
\end{itemize}
All of these nodes reside in the main cluster of our graph so we ask
\begin{itemize}
\item How are nodes in the main cluster connected to the Tier 1 and 2 standards? 
\item Can we improve this connectivity with the nodes we have identified as important?
\end{itemize}

\begin{figure}
\caption{\label{figure:distanceDist}Plot of the minimum distance to a standard.  We show a line at the average path length for the main cluster $\ell=5.508$.  This is the cluster where all of the standards nodes reside.}
\includegraphics[width=0.5\textwidth]{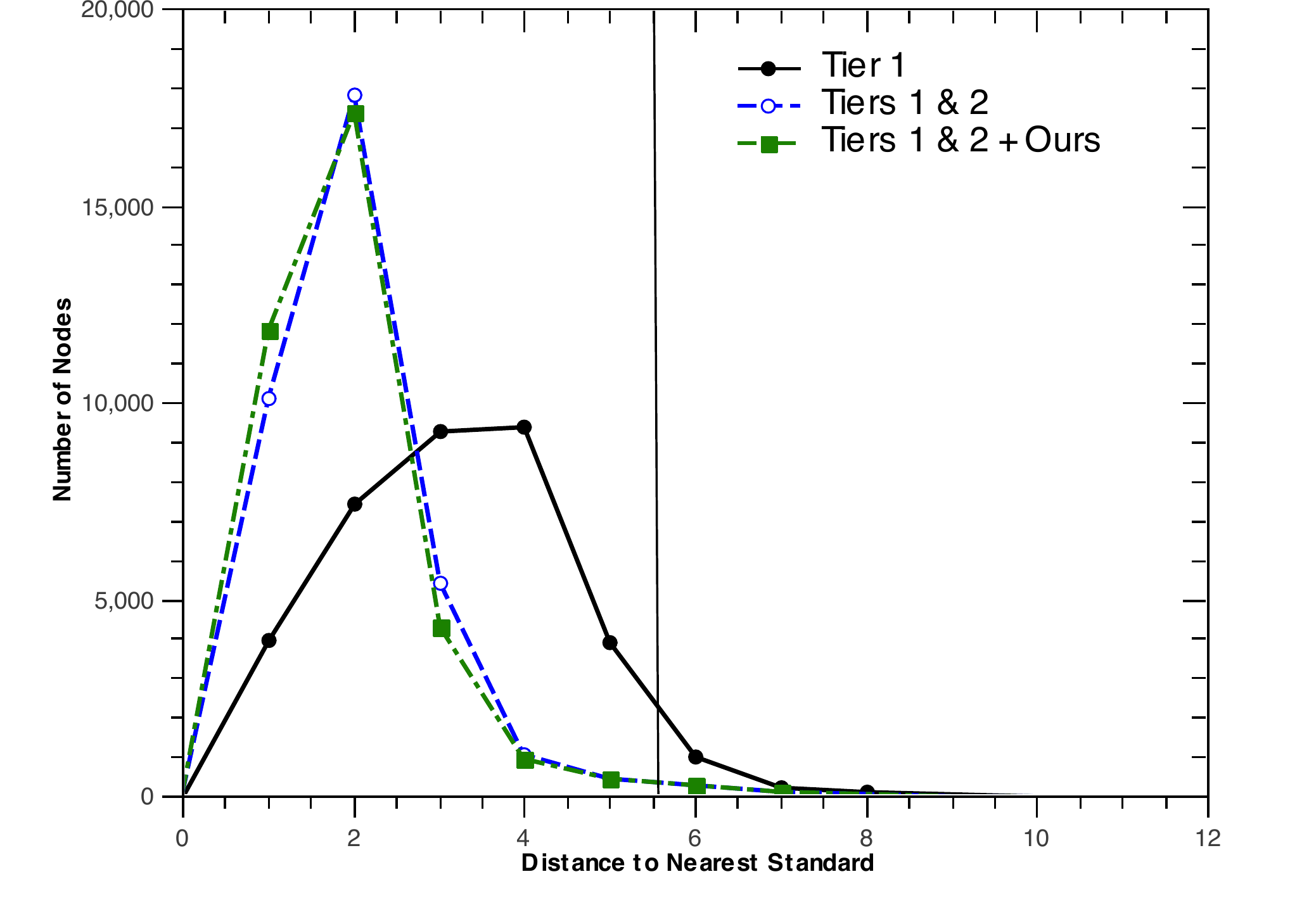}
\end{figure}

The simplest measure of graph connectivity is the mean distance between nodes.  The distance between any two connected nodes is the minimum number of edges separating the nodes including all possible paths between the nodes.  Here we are not interested in the distance between arbitrary nodes but are interested in the distance between any node and a Tier 1 or 2 standard node.  In Figure \ref{figure:distanceDist} we show several histograms of distance from nodes to the nearest standards node for several cases.  

In this plot, nodes with distance zero are the standards themselves.  Considering just the Tier 1 standards, the distribution is rather broad and peaks at a distance of 5 nodes and extends to 10 nodes.  We note that this peak is near $\ell = 5.508$, the average path length in the main cluster.  The main cluster is a ``small world'' graph, so it has tight clusters within that can be used to increase connectivity, like the hubs of an airline network. Adding in the large number of Tier 2 standards dramatically tightens up the distribution with the peak now at 2 the distribution now extends to 8 nodes.  As we saw earlier, many of these Tier 2 standards have high degree and can function as hubs.  Adding the nodes corresponding to our proposed list of nodes tightens the distribution up further, enhancing the connectivity to standards level nodes.  We comment that there is a noticeable improvement from adding our seven proposed nodes, a surprisingly large improvement given the small number of added nodes. 

\section{Conclusion}
In this project, we created an undirected graph from the {\tt REACTION} and {\tt MONITOR} strings from datasets in the EXFOR database.  This graph is a large, nearly scale-free network composed of disconnected clusters.  The largest clusters have a ``small-world'' character.  Our graph is in many ways typical for real world graphs.

With our graph, we identify what reactions and quantities the nuclear science community views as important enough to directly measure or measure relative to.  We do this in a relatively objective fashion.  Clearly the various standards projects in Refs. \cite{Standards,Atlas,IAEAMedical,IRDFF} have a good handle on what is important.  Also, the  clustering coefficients in Table \ref{table:topByClusterCoeff} demonstrate how connected the CIELO nodes are. However, it is clear from the analysis of our graph that the following reaction/quantities have out-sized importance and are not considered in any standards effort: 
\begin{itemize}
\item Aluminum reaction/quantities:
\begin{itemize}
\item n+$^{27}$Al: the ($n,p+^{27}$Mg) cross section
\item p+$^{27}$Al: the ($p,n+3p$) cross section and the $^{22}$Na and $^{24}$Na production cross sections
\item $^{12}$C+$^{27}$Al: the $^{24}$Na production cross section
\end{itemize}
\item Molybdinum also a very important structural material:
\begin{itemize}
\item p+$^{nat}$Mo: the $^{96}$Tc production cross section
\item $\alpha$ + $^{nat}$Mo: the $^{97}$Ru production cross section
\end{itemize}
\end{itemize}
We recommend that at the very least that $^{27}$Al and all of the Mo isotopes be considered as a target material in either a follow-on CIELO or IRDFF project.  In addition, a standards level study of fission product yields of the major actinides as suggested in the discussions at the recent Working Party on Evaluation Cooperation Subgroup 37 meeting \cite{WPECSG37} would improve the connectivity of all fission product yield data.

\section*{Acknowledgments}
We want to thank M. Herman (BNL) and J. Fritz (St. Joseph's College) for their support of this project and acknowledge the useful discussions with N. Otsuka (IAEA), A. Carlson (NIST), A. Plompen (IRMM) and R. Capote (IAEA).  The work at Brookhaven National Laboratory was sponsored by the Office of Nuclear Physics, Office of Science of the U.S. Department of Energy under Contract No. DE-AC02-98CH10886 with Brookhaven Science Associates, LLC.  This project was supported in part by the U.S. Department of Energy, Office of Science, Office of Workforce Development for Teachers and Scientists (WDTS) under the Science Undergraduate Laboratory Internships Program (SULI).


\end{document}